\documentclass[twocolumn,showpacs,preprintnumbers,amsmath,amssymb,pre,cite]{revtex4}

\usepackage{graphicx}% Include figure files
\usepackage{dcolumn}% Align table columns on decimal point
\usepackage{bm}% bold math

\usepackage{color}

\providecommand{\vc}[1]{\mathbf{#1}}

\newcommand{\uij}{\vc{u}_{ij}}
\newcommand{\rij}{\vc{r}_{ij}}
\newcommand{\uparl}{u_\parallel}
\newcommand{\uperp}{u_\perp}

\newcommand{\dm}{\mathcal{M}}

\usepackage{amssymb}

\begin{document}

\title[Jamming]{Jamming of Soft Particles:
Geometry, Mechanics, Scaling and Isostaticity}

\author{M van Hecke}

\address{Kamerlingh Onnes Laboratory, Leiden University, PO box
9504,\\
2300 RA Leiden, The Netherlands.\\
mvhecke@physics.leidenuniv.nl}

\begin{abstract}

Amorphous materials as diverse as foams, emulsions, colloidal
suspensions and granular media can {\em jam} into a rigid,
disordered state where they withstand finite shear stresses before
yielding. Here we review the current understanding of the
transition to jamming and the nature of the jammed state for
disordered packings of particles that act through repulsive
contact interactions and are at zero temperature and zero shear
stress. We first discuss the breakdown of affine assumptions that
underlies the rich mechanics near jamming. We then extensively
discuss jamming of frictionless soft spheres. At the jamming
point, these systems are marginally stable (isostatic) in the
sense of constraint counting, and many geometric and mechanical
properties scale with distance to this jamming point. Finally we
discuss current explorations of jamming of frictional and
non-spherical (ellipsoidal) particles. Both friction and
asphericity tune the contact number at jamming away from the
isostatic limit, but in opposite directions. This allows one to
disentangle distance to jamming and distance to isostaticity. The
picture that emerges is that most quantities are governed by the
contact number and scale with distance to isostaticity, while the
contact number itself scales with distance to jamming.

\end{abstract}


 \maketitle

\section{Introduction}

\begin{figure*}\hspace{6mm}
\includegraphics[width=13.4cm,viewport=1 1 600 330,clip]{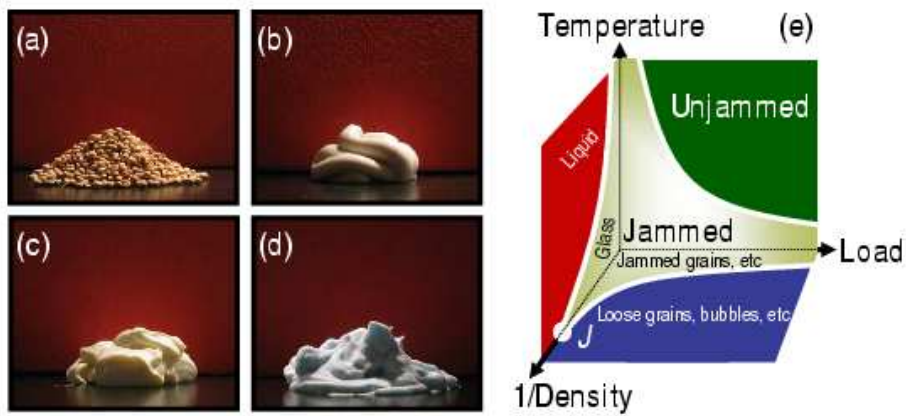} \caption{(a-d) Examples of everyday
disordered media in a jammed state. (a) Granular media, consisting
of solid grains in gas or vacuum. (b) Toothpaste, a dense packing
of (colloidal) particles in fluid. (c) Mayonnaise, an emulsion
consisting of a dense packing of (oil) droplets in an immiscible
fluid. (d) Shaving foam, a dense packing of gas bubbles in fluid.
(e) Jamming diagram proposed by Liu, Nagel and co-workers
\cite{jamnote,epitome}. The diagram illustrates that many
disordered materials are in a jammed state for low temperature,
low load and large density, but can yield and become unjammed when
these parameters are varied. In this review we will focus on the
zero temperature, zero load axis. For frictionless soft spheres,
there is a well defined jamming transition indicated by point
``J'' on the inverse density axis, which exhibits similarities to
an (unusual) critical phase transition. }\label{fig:jam}
\end{figure*}

Jamming governs the transition to rigidity of disordered matter.
Foams, emulsions, colloidal suspensions, pastes, granular media
and glasses can {\em jam} in rigid, disordered states in which
they respond essentially elastically to small applied shear
stresses (Fig.~\ref{fig:jam}a-d). However, they can also easily be
made to yield (unjam) and flow by tuning various control
parameters.

The transition from the freely flowing to the jammed state, the
jamming transition, can be induced by varying thermodynamic
variables, such as temperature or density, but also mechanical
variables such as the stress applied to the sample: colloidal
suspensions become colloidal glasses as the density is increased
near random close packing, flowing foams become static as the
shear stress is decreased below the yield stress, and supercooled
liquids form glasses as the temperature is lowered below the glass
transition temperature. In 1998 Liu and Nagel presented their
provocative jamming phase diagram (Fig.~\ref{fig:jam}e), and
proposed to probe the connections between various transitions to
rigidity \cite{jamnote}.

This review provides an overview of the current (partial) answers
to the following two questions: What is the nature of the jammed
state? What is the nature of the jamming transition? We focus on
jammed model systems at zero temperature and zero shear --- models
for non-brownian emulsions, foams and granular media rather than
colloidal and molecular glasses --- and review the geometrical and
mechanical properties of these systems as a function of the
distance to jamming.

In view of the very rapid developments in the field, the paper
focuses on the basic jamming scenarios, which arises in (weakly)
compressed systems of soft particles interacting through repulsive
contact forces at zero temperature and zero shear. The picture
that has emerged for the jamming transition in these systems is
sufficiently complete to warrant an overview article and, in
addition, provides a starting point for work on a wider range of
phenomena, such as occurring in attractive systems
\cite{coreyattract}, systems below jamming \cite{below}, the flow
of disordered media near jamming
\cite{olsson,langlois,katgert,head,hatano08}, jamming of systems
at finite temperature \cite{berthier,vestige} and experiments
\cite{behringerdz,dauchot,durian}.

In this review the focus is on jamming of frictionless spheres,
frictional spheres, and frictionless ellipsoids --- soft
(deformable) particles which interact through repulsive contact
forces. The distance to jamming of all these systems is set by the
amount of deformation of the particles, which can be controlled by
the applied pressure or enforced packing fraction. These systems
lose rigidity when the deformations vanish, or equivalently, when
the confining pressure reaches zero. As we will see, these
seemingly simple systems exhibit rich and beautiful behavior,
where geometry and mechanical response are intricately linked.

The contact number, $z$, defined as the average number of contacts
per particle, plays a crucial role for these systems. There is a
minimal value of $z$ below which the system loses rigidity: when
the contact number is too small, there are collective particle
motions, so-called floppy modes, that (in lowest order) do not
cost elastic energy. By a constraint counting argument one can
establish a precise value for the minimum value of $z$ where the
system does not generically allows floppy deformations --- this is
the isostatic contact number $z_{\rm iso}$. As we will see, a host
of mechanical and geometrical properties of jammed systems scale
with distance to the isostatic point.

The crucial, and at first glance very puzzling point, is that
while frictionless spheres reach isostaticity at the jamming
point, frictional spheres are generally hyperstatic ($z>z_{\rm
iso}$) at jamming, while frictionless ellipsoids are hypostatic
($z<z_{\rm iso}$) at jamming. As we will see, the relations
between contact numbers, floppy modes, rigidity and jamming are
subtle.

Truly new and surprising physics emerges near jamming in systems
as seemingly simple as disordered packings of frictionless,
deformable particles \cite{epitome}. We first discuss the
breakdown of affine assumptions that underlies the rich physics of
jamming in section~\ref{sec:motiv}. We give an overview of the
main characteristics of the jamming transition for soft
frictionless spheres in section~\ref{sec:slippery}. Both friction
and asphericity lead to new physics, as here the jamming
transition and isostaticity decouple. Jamming of frictional soft
spheres is discussed in section~\ref{sec:fric}, and jamming of
frictionless soft ellipsoids in section~\ref{sec:ellipsoid}.
Finally, in section~\ref{sec:disc} we sketch a number of open
problems.

\section{Motivation: Mechanics of Disordered
Matter}\label{sec:motiv}

The crucial question one faces when attempting to describe the
mechanics of materials such as foams, emulsions or granular media,
is how to deal with disorder. The simplest approach is to ignore
disorder altogether, and attempt to gain insight based on models
for ordered, ``crystalline'' packings. A related approach,
effective medium theory, does not strictly require ordered
packings, but assumes that local deformations and forces scale
similarly as global deformations and stresses. As we will see in
section~\ref{rigsec}, major discrepancies arise when these
approaches are confronted with (numerical) experiments on
disordered systems. This is because the response of disordered
packings becomes increasingly non-affine near jamming
(section~\ref{sec:beyond}).

\subsection{Failure of Affine Approaches} \label{rigsec}

\subsubsection{Foams and Emulsions}

Some of the earliest studies that consider the question of
rigidity of packings of particles concern the loss of rigidity in
foams and emulsions with increasing wetness. Foams are dispersions
of gas bubbles in liquid, stabilized by surfactant, and the gas
fraction $\phi$ plays a crucial role for the structure and
rigidity of a foam. The interactions between bubbles are repulsive
and viscous, and static foams are similar to the frictionless soft
spheres discussed in section~\ref{sec:slippery}. In real foams,
gravity (which causes drainage) and gas diffusion (which causes
coarsening) play a role, but we will ignore these.

\begin{figure}[t]
\resizebox{8.6cm}{!}{\includegraphics[viewport=30 30 480
170,clip]{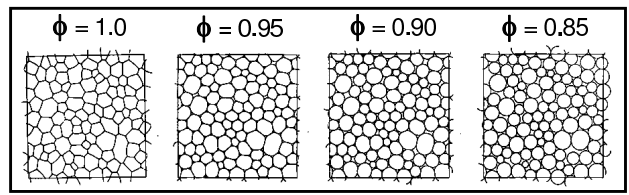}} \caption{Simulated foam for increasing
wetness, approaching unjamming for $\phi \downarrow 0.84$ (adapted
from \cite{bolton} --- Copyright by the American Physical
Society).}\label{fig:bolton}
\end{figure}

The unjamming scenario for foams is as follows. When the gas
fraction approaches one, the foam is called dry. Application of
deformations causes the liquid films to be stretched, and the
increase in surface area then provides a restoring force: dry
foams are jammed. When the gas fraction is lowered and the foam
becomes wetter, the gas bubbles become increasingly spherical, and
the foam loses rigidity for some critical gas fraction $\phi_c$
where the bubbles lose contact (Fig.~\ref{fig:bolton}). The
unjamming transition is thus governed by the gas fraction, which
typically is seen as a material parameter. For emulsions,
consisting of droplets of one fluid dispersed in a second fluid
and stabilized by a surfactant, the same scenario arises.

Analytical calculations are feasible for ordered packings, because
one only needs to consider a single particle and its neighbors to
capture the packing geometry and mechanical response of the foam
--- due to the periodic nature of the packing, the
response of the material is affine. The affine assumption
basically states that locally, particles follow the globally
applied deformation field
--- as if the particles are pinned to an elastically deforming
sheet. More precisely, the strict definition of affine
transformations states that three collinear particles remain
collinear and that the ratio of their distances is preserved, and
affine transformations are, apart from rotations and translations,
composed of uniform shear and compression or dilatation.

Packings of monodisperse bubbles in a two-dimensional hexagonal
lattice (``liquid honeycomb'' \cite{kraynik}) deform affinely. The
bubbles lose contact at the critical density $\phi_c$ equal to
$\frac{\pi}{2\sqrt{3}} \approx0.9069$, and ordered foam packings
are jammed for larger densities \cite{kraynik,princen83}. When for
such a model foam $\phi$ is lowered towards $\phi_c$, the yield
stress and shear modulus remain finite, and jump to zero precisely
at $\phi_c $\cite{kraynik,princen83}. The contact number (average
number of contacting neighbors per bubble) remains constant at 6
in the jammed regime. Similar results can be obtained for
three-dimensional ordered foams, where $\phi_c$ is given by the
packing density of the HCP lattice $\frac{\pi}{3\sqrt{2}} \approx
0.7405$.

Early measurements for polydisperse emulsions by Princen and Kiss
in 1985 \cite{princenIII} found a shear modulus which varied
substantially with $\phi$. Even though no data was presented for
$\phi$ less than 0.75 and the fit only included points for which
$\phi \ge 0.8$, the shear modulus was fitted as  $G \sim
\phi^{1/3} (\phi-\phi_c)$, where $\phi_c \approx 0.71$, and thus
appeared to vanish at a critical density below the value predicted
for ordered lattices \cite{princenIII}.

The fact that the critical packing density for ordered systems is
higher than that for disordered systems may not be a surprise,
given that at the jamming threshold, the particles are undeformed
spheres, and  it is well known that ordered sphere packings are
denser than irregular ones \cite{davescience}. However, the
differences between the variation of the moduli and yield strength
with distance to the rigidity threshold predicted for ordered
packings and measured for disordered emulsions strongly indicates
that one has to go beyond models of ordered packings.

\subsubsection{Effective Medium Theory for Granular Media} For
granular media an important question has been to predict the bulk
elasticity, and Makse and co-workers have carried out extensive
studies of the variation of the elastic moduli and sound
propagation speed with pressure in granular media from the
perspective of effective medium theory \cite{makseprl1999,
maksephysb2000,maksepre2004}.

Effective medium theory (EMT) basically assumes that: {\em{(i)}}
Macroscopic, averaged quantities can be obtained by a simple
coarse graining procedure over the individual contacts.
{\em{(ii)}} The effect of global forcing, e.g., imposing a
deformation, trivially translates to changes in the local
contacts. This second assumption is the ``affine assumption'', and
this will be the crucial assumption that breaks down near jamming.

Makse {\em et al.}~studied the breakdown of effective medium
theory in the context of granular media. Assuming a Hertzian
interaction between spherical grains \cite{Hertzian}, the contact
force $f$ scales with the overlap $\delta$ between particles as
$f\sim \delta^{3/2}$. As a result, the stiffness of these contacts
then scales as $\partial_{\delta} f \sim \delta^{1/2}$. Since, in
good approximation, the pressure $P\sim f$, one obtains that the
stiffness of the individual contacts scales as $P^{1/3}$. EMT then
predicts that the elastic bulk modulus $K$ and shear modulus $G$
scale as the stiffness of the contacts: $K\sim G \sim P^{1/3}$,
and that the sound velocities scales as $P^{1/6}$
\cite{ellakpre2005,makseprl1999,maksephysb2000,maksepre2004}. In
particular, the ratio $G/K$ should be independent of pressure.

From a range of simulations Makse {\em et al.}~concluded that the
affine assumption works well for the compression modulus provided
that the change in contact number with $P$ is taken into account,
but fails for the shear modulus --- and suggested that this is due
to the non-affine nature of the deformations
\cite{makseprl1999,maksephysb2000,maksepre2004}. We will discuss
this issue at length in section \ref{sec:slippery}.

\begin{figure}[t]
\hfill\resizebox{8.3cm}{!}{\includegraphics[viewport=30 20 290
240,clip]{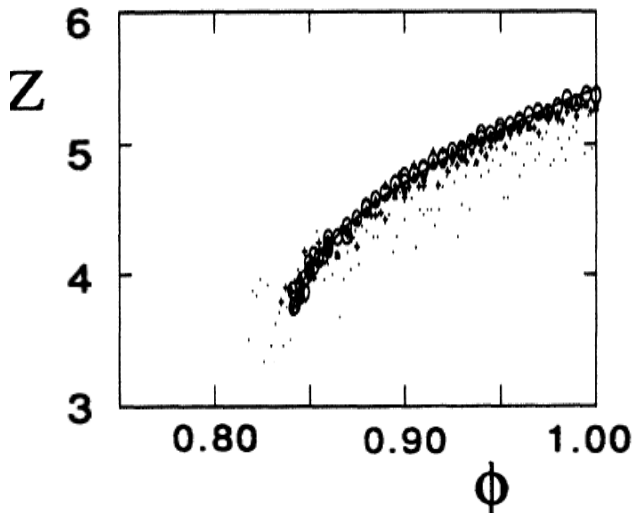}}\caption{Square root scaling of contact
number $z$ with $\phi-\phi_c$ observed in the Durian bubble model
(adapted from \cite{durianbubble} --- Copyright by the American
Physical Society).}
    \label{fig:durian}
\end{figure}

\subsection{Beyond Affine Approaches}\label{sec:beyond}

In a seminal paper in 1990, Bolton and Weaire asked how a {\em
disordered} foam loses rigidity when its gas fraction is decreased
\cite{bolton}. They probed this question by simulations of a
two-dimensional polydisperse foam, consisting of approximately
hundred bubbles, as a function of $\phi$ (Fig.~\ref{fig:bolton}).
Their model captures the essential surface tension driven
structure of foams and predates the now widely used ``surface
evolver'' code for foams \cite{surfaceevolver}.

The following crucial observations are made: {\em{(i)}} The
critical density is around 0.84, which is identified as the random
close packing density in two dimensions --- here the yield stress
appears to vanish smoothly. {\em{(ii)}} The contact number $z$
smoothly decreases with $\phi$. At $\phi=1$ the contact number
equals six. This can be understood by combining Euler's theorem
which relates the number of vertices, faces and edges in tilings
with Plateau's rule that for a two-dimensional dry foam in
equilibrium, three films (faces) meet in one point (vertex). When
$\phi \rightarrow \phi_c$, the contact number appears to reach the
marginal value, four. {\em{(iii)}} The shear modulus decreases
with $\phi$ and appears to smoothly go to zero at $\phi = \phi_c$
(unfortunately the authors do not comment on the bulk modulus).

In related work on the so-called bubble model developed for wet
foams in 1995, Durian reached similar conclusions for
two-dimensional model foams, and moreover found that the contact
number indeed approaches $4(=2d)$ near jamming, and observed the
non-trivial square root scaling of $z-4$ with excess density for
the first time (Fig~\ref{fig:durian}). All these findings are
consistent with what is found in closely related models of
frictionless soft spheres near jamming, as discussed in sections
\ref{sec:slippery}.

Experimentally, measurements of the shear modulus and osmotic
pressure of compressed three-dimensional monodisperse but
disordered emulsions found similar behavior for the loss of
rigidity \cite{mason95,mason96,mason97}. The shear modulus, (when
scaled appropriately with the Laplace pressure, which sets the
local ``stiffness'' of the droplets) grows continuously with
$\phi$ and vanishes at $\phi_c \approx 0.635$, corresponding to
random close packing in three dimensions. The osmotic pressure
exhibits very similar scaling, implying that the bulk modulus
(being proportional to the derivative of the pressure with respect
to $\phi$) scales differently from the shear modulus --- the
difference between shear and bulk modulus is another hallmark of
jamming of frictionless spheres.

\begin{figure}[t]
\hfill
\includegraphics[width=8.6cm,viewport=40 30 530 520,clip]{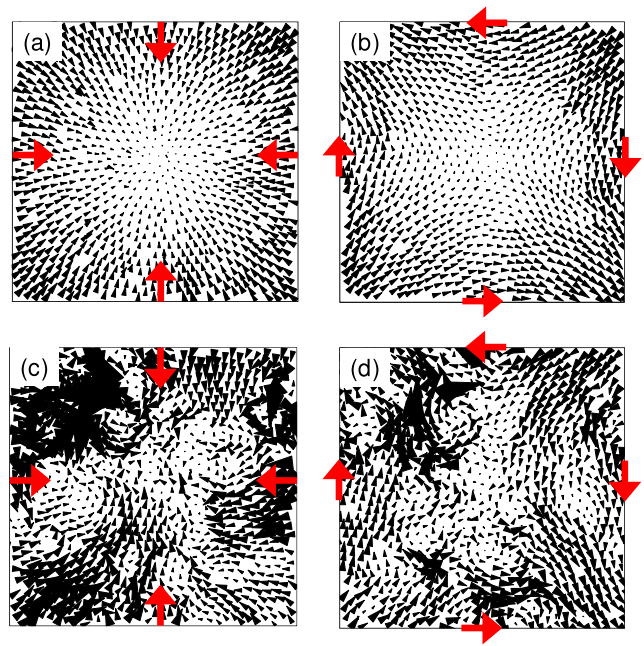}
\caption{Deformation fields of packings of 1000 frictionless
particles under compression (a,c) and shear (b,d) as indicated by
the red arrows. The packings in the top row (a,b) are strongly
jammed (contact number $z=5.87$), while the packings in the bottom
row (c,d) are close to the jamming point --- their contact number
is $4.09$, while the jamming transition occurs for $z=4$ in this
case. Clearly, the deformation field becomes increasingly
non-affine when the jamming point is approached (adapted from
\cite{ellenbroek2006,wouterlinlang} --- Copyright by the American
Physical Society).}\label{fig:affine}
\end{figure}

There is thus a wealth of simulational and experimental evidence
that invalidates simple predictions for the rigidity of disordered
media based on our intuition for ordered packings. The crucial
ingredient that is missing is the non-affine nature of the
deformations of disordered packings (Fig.~\ref{fig:affine}). There
is no simple way to estimate the particles motion and deformations
in disordered systems, and one needs to resort to (numerical)
experiments. Jamming can be seen as the avenue that connects the
results of such experiments. Jamming aims at capturing the
mechanical and geometric properties of disordered systems,
building on two insights: first, that the non-affine character
becomes large near the jamming transition, and second, that
disorder and non-affinity are not weak perturbations away from the
ordered, affine case, but may lead to completely new physics
\cite{ellakpre2005,mason95,granulence,tanguy,tanguy04,
lemaitre06,maloneyPRL06}.

\section{Jamming of Soft Frictionless Spheres}\label{sec:slippery}

\begin{figure}
\hfill\includegraphics[width=8.6cm,viewport=90 40 600
280,clip]{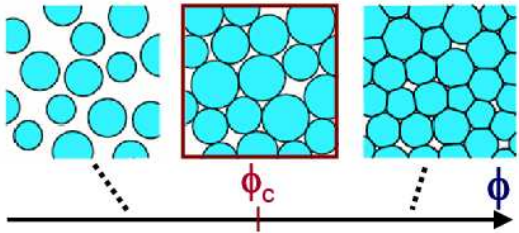} \caption{States of soft frictionless spheres
as function of packing density $\phi$, below, at, and above the
critical density $\phi_c$. Left: Unjammed system at a density
below the critical density --- pressure is zero and there are no
contacts. Middle: Marginally rigid system consisting of undeformed
frictionless spheres just touching. The system is at the jamming
transition (point J), has vanishing pressure, critical density and
$2d$ contacts per particle, where $d$ is the dimension. Right:
Jammed system for finite pressure and density above
$\phi_c$.}\label{fig:J}
\end{figure}

Over the last decade, tremendous progress has been made in our
understanding of what might be considered the ``Ising model'' for
jamming: static packings of soft, frictionless spheres that act
through purely repulsive contact forces. In this model,
temperature, gravity and shear are set to zero. The beauty of such
systems is that they allow for a precise study of a jamming
transition. As we will see in sections \ref{sec:fric} and
\ref{sec:ellipsoid}, caution should be applied when applying the
results for soft frictionless spheres to frictional and/or
non-spherical particles.

From a theoretical point of view, packings of soft frictionless
spheres are ideal for three reasons. First, they exhibit a well
defined jamming point: For positive $P$ the system is jammed, as
it exhibits a finite shear modulus and a finite yield stress
\cite{epitome}, while at zero pressure the systems loses rigidity.
Hence, the (un)jamming transition occurs when the pressure $P$
approaches zero, or, geometrically, when the deformations of the
particles  vanish. The zero pressure, zero shear, zero temperature
point in the jamming phase diagram is referred to as ``point J''
(Fig.~\ref{fig:jam}e and \ref{fig:J}). In this review, point J
will only refer to soft frictionless spheres and not to jamming
transitions of other types of particles. Second, at point J the
contact number approaches the so-called isostatic value, and the
system is marginally stable. The system's mechanical and
geometrical properties are rich and peculiar here. For large
systems the critical packing density, $\phi_c$, approaches values
usually associated with random close packing. Third, the
mechanical and geometrical properties of jammed systems at finite
pressure, or equivalently, $\phi-\phi_c>0$, exhibit non-trivial
power law scalings as a function $\Delta \phi:= \phi-\phi_c$ or,
similarly, as function of the pressure, $P$.

In this section we address the special nature of point J and
discuss the scaling of the mechanical and geometrical properties
for jammed systems near point J. We start in section
\ref{sec:defmod} by a brief discussion of a few common contact
laws and various numerical protocols used to generate jammed
packings. We then present evidence that the jamming transition of
frictionless spheres is sharp and discuss the relevant control
parameters in section \ref{sec:sharp}. In section \ref{sec:geomJ}
we discuss the special geometrical features of systems at point J,
as probed by the contact number and pair correlation function.
Away from point J the contact number exhibits non-trivial scaling,
which appears to be closely related to the pair correlation
function at point J, as discussed in section \ref{sec:awayJ}. Many
features of systems near point J can be probed in linear response,
and these are discussed at length in section \ref{sec:linresp} ---
these include the density of states (\ref{secdos}), diverging
length and time scales (\ref{secscale}), elastic moduli
(\ref{secmodul}) and non-affine displacements (\ref{secnonaf}). We
close this section by a comparison of effective medium theory,
rigidity percolation and jamming, highlighting the unique nature
of jamming near point J (\ref{emtrp}).

\subsection{Definition of the Model}\label{sec:defmod}

At the (un)jamming transition soft particles are undeformed, and
the distance to jamming depends on the amount of deformation.
Rigid particles are therefore always at the jamming transition,
and soft particles are necessary to vary the distance to point J.
Deformable frictionless spheres interact through purely repulsive
body centered forces, which can be written as a function of the
amount of virtual overlap between two particles in contact.
Denoting the radii of particles in contact as $R_i$ and $R_j$ and
the center-to-center distance as $r_{ij}$, it is convenient to
define a dimensionless overlap parameter $\delta_{ij}$ as
\begin{equation}\label{deltadef}
\delta_{ij} := 1 - \frac{r_{ij}}{R_i+R_j}~,
\end{equation}
so that particles are in contact only if $\delta_{ij}\ge 0$. We
limit ourselves here to interaction potentials of the form:
\begin{eqnarray}\label{eqs:interact}
V_{ij} &= \epsilon_{ij}~ \delta_{ij}^\alpha \hspace{10mm} &\delta_{ij} \ge 0~, \\
V_{ij} &= 0~~~~~~~~~~~~~~  &\delta_{ij} \le 0~.
\end{eqnarray}
By varying the exponent, $\alpha$, one can probe the nature and
robustness of the various scaling laws discussed below. For
harmonic interactions, $\alpha = 2$ and $\epsilon_{ij}$ sets the
spring constant of the contacts. Hertzian interactions between
three-dimensional spheres, where contacts are stiffer as they are
more compressed, correspond to $\alpha=5/2$ \footnote{When one
strictly follows Hertz Law, one finds that $\epsilon_{ij}$ depends
on the radii $R_i$ and $R_j$
--- but often $\epsilon_{ij}$ is simply takes as a constant, and
for typical polydispersities, the effect of this for statistical
properties of packings is likely small \cite{wouterlinlang}.}.
O'Hern {\em et al}~have  also studied the ``Hernian'' interaction
($\alpha=3/2$), which corresponds to contacts that become
progressively weaker when compressed \cite{epitome}.

Once the contact laws are given, one can generate packings by
various different protocols, of which MD (Molecular Dynamics)
\cite{makseprl1999,maksephysb2000,maksepre2004,ellakpre2005} and
conjugate gradient \cite{epitome} are the most commonly used
\footnote{For undeformable particles, the Lubachevsky-Stillinger
algorithm can be used}. In MD simulations one typically starts
simulations with a loose gas of particles, which are incrementally
compressed, either by shrinking their container or by inflating
their radii. Supplementing the contact laws with dissipation
(inelastic collisions, viscous drag with a virtual background
fluid, etc) the systems ``cools'' and eventually one obtains a
stationary jammed state. While straightforward, one might worry
that statistical properties of packings obtained by such procedure
depend on aspects of the procedure itself
--- for frictional packings, this is certainly the case
\cite{kasahara}.

For frictionless particles, the interactions are conservative, and
one can exploit the fact that stable packings correspond to minima
of the elastic energy. Packings can then be created by starting
from a completely random configuration and then bringing the
system to the nearest minimum of the potential energy. When the
energy at this minimum is finite, the packing is at finite
pressure, and this procedure is purported to sample the phase
space of allowed packings flatly
\cite{epitome,discussionafterwards}. An effective algorithm to
find such minima is known as the ``conjugate gradient technique''
\cite{conjugate}. For frictionless systems, we are not aware of
significant differences between packings obtained by MD and by
this method \footnote{It is an open question whether history never
plays a role for frictionless spheres --- for example, one may
imagine that by repeated decompression and recompression,
different ensembles of packings could be accessed.}.

Finally it should be noted that to avoid crystallization,
two-dimensional packings are usually made polydisperse, and a
popular choice are bidisperse packings where particles of radii 1
and 1.4 are mixed in equal amounts \cite{epitome,ellenbroek2006}.
In three dimensions, this is not necessary as monodisperse spheres
then do not appear to order or crystallize for typically employed
numerical packing generation techniques.

\subsection{Evidence for Sharp Transition}\label{sec:sharp}

The seminal work of O'Hern {\em et al.}~\cite{epitome,ohern02} has
laid the groundwork for much of what we understand about jamming
of frictionless soft spheres. These authors begin by carefully
establishing that frictionless soft spheres exhibit a sharp
jamming transition. First, it was found that when a jammed packing
is decompressed, the pressure, the bulk modulus and the shear
modulus vanish at the same critical density $\phi_c$. For finite
systems, the value of $\phi_c$ varies from system to system. For
systems of 1000 particles the width of the distribution of
$\phi_c$, $W$, still corresponds to 0.4\%, and must therefore not
be ignored. Second, it was shown that the width, $W$, vanishes
with the number of particles $N$ as $W \sim N^{-1/2}$
--- independent of dimension, interaction potential or polydispersity.
In addition, the location of the peak of the distribution of
$\phi_c$,  $\phi_0$, also scales with $N$:
 $\phi_0 - \phi^* = (0.12 \pm 0.03) N^{-1/\nu
d}$. Here $d$ is the dimensionality, $\nu=0.71 \pm 0.08$ and
$\phi^*$ approaches $0.639 \pm 0.001$ for three-dimensional
monodisperse systems.

These various scaling laws suggest that for frictionless spheres
the jamming transition is sharp in the limit of large systems.
This jamming point is referred to as point $J$ (see
Fig.~\ref{fig:jam}e and ~\ref{fig:J}). At the jamming point, the
packings consist of perfectly spherical (i.e., undeformed) spheres
which just touch (Fig.~\ref{fig:J}). The packing fraction for
large systems, $\phi^*$, reaches values which have been associated
with random close packing (RCP) \cite{bolton,epitome} --- ($\sim
0.84$ in two dimensions, $\sim 0.64$ in three dimensions). It
should be noted that the RCP concept itself is controversial
\cite{rcpdiscussion}.

{\em Control Parameters ---} As we will see, the properties of
packings of soft slippery balls are controlled by their distance
to point J. What is a good control parameter for jamming at point
J? The spread in critical density for finite systems indicates
that one should not use the density, but only the excess density
$\Delta \phi := \phi-\phi_c$ as control parameter. In other words,
fixing the volume is not the same as fixing the pressure for
finite systems.

The disadvantage of using the excess density is that it requires
deflating packings to first obtain $\phi_c$ \cite{epitome}. This
extra step is not necessary when $P$ is used as control parameter,
since the jamming point corresponds to $P=0$ --- no matter what
the system size or $\phi_c$ is of a given system. While we believe
it is much simpler to deal with fixed pressure than with fixed
volume, a disadvantage of $P$ is that its relation to $\Delta
\phi$ is interaction dependent: the use of the excess density
stresses the geometric nature of the jamming transition at point
J.

We suggest that the average overlap $\left<\delta\right>$ is the
simplest control parameter  --- even though its use is not common.
First, $\left<\delta\right>$ is geometric and interaction
independent and reaches zero at jamming, also for finite systems.
Moreover, for finite systems $\left<\delta\right>$ still controls
the pressure and will be very close to $ \Delta \phi$. Of course,
in infinite systems, control parameters like the pressure $P$, the
average particle overlap $\left<\delta\right>$ and the density
$\phi$ are directly linked
--- for interactions of the form Eq.~(\ref{eqs:interact}), $P \sim
\delta^{\alpha-1} \sim (\Delta \phi)^{\alpha-1}$. Below, we will
use a combination of all these control parameters, reflecting the
different choices currently made in the field.

\subsection{Geometry at Point J}\label{sec:geomJ}

At point J, the system's packing geometry is highly non-trivial.
First, systems at point J are isostatic \cite{alexander}: the
average number of contacts per particle is sharply defined and
equals the minimum required for stability
\cite{epitome,moukarzel,tchachenko}. Second, near jamming $g(r)$
diverges when $r \downarrow 1$ (for particles of radius 1)
\cite{silbertPRE06,silbert2002,donevgr}.

{\em Isostaticity ---} The fact that the contact number at point J
attains a sharply defined value has been argued to follow directly
from counting the degrees of freedom and constraints
\cite{moukarzel,tchachenko}. We discuss such counting arguments in
detail in Appendix \ref{sec:count}, but give here the gist of the
argument for frictionless spheres.

Suppose we have a packing of $N$ soft spheres in $d$ dimensions,
and that the contact number, the average number of contacts at a
particle, equals $z$
--- the total number of contacts equals $zN/2$, since each contact
is shared by two particles. First, the resulting packing should
not have any floppy modes, deformation modes that cost zero energy
in lowest order. As we discuss in Appendix \ref{sec:count}, this
is equivalent to requiring that the $N z/2$ contact forces balance
on all grains, which yields $dN$ constraints on $N z/2$ force
degrees of freedom: hence $z \ge 2d$. The minimum value of $z$
required is referred to as the isostatic value $z_{\rm iso}$: for
frictionless spheres, $z_{\rm iso} = 2d$.

Second, at point J, since the particles are undeformed: the
distance between contacting particles has to be precisely equal to
the sum of their radii. This yields $N z/2$ constraints for the
$dN$ positional degrees of freedom: therefore, one only expects
generic solutions at jamming when $z \le 2d$.

Combining these two inequalities then yields that the contact
number $z_c$ at the jamming point for soft frictionless disks
generically will attain the isostatic value: $z_c=z_{\rm iso}=2d$
\cite{epitome,moukarzel,tchachenko}. As we will see below, such
counting arguments should be regarded with caution, since they do
not provide a correct estimate for the contact number at jamming
of frictionless ellipsoidal particles
\cite{donev2007,coreyellipses,zz2009}.

Numerically, it is far from trivial to obtain convincing evidence
for the approach of the contact number to the isostatic value.
Apart from corrections due to finite system sizes and finite
pressures, a subtle issue is how to deal with rattlers, particles
that do not have any contacts with substantial forces, but still
arise in a typical simulation. These particles have low
coordination number and their overlap with other particles is set
by the numerical precision
--- these particles do not contribute to rigidity.
For low pressures, they can easily make up 5\% of the particles.
An accurate estimate of the contact number than requires one to
ignore these particles and the corresponding ``numerical''
contacts \cite{epitome,shundjak2007}.

\begin{figure}
\begin{center}
\includegraphics[width=10.6cm,viewport=40 20 420 217,clip]{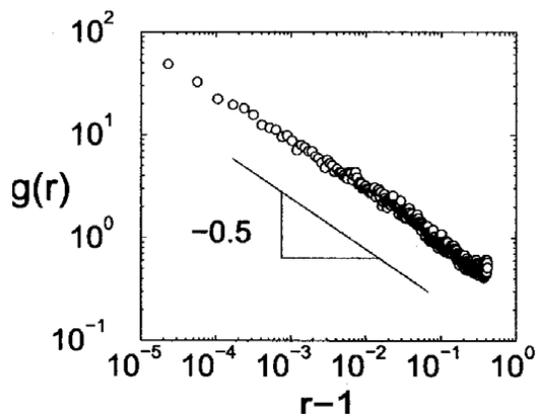}
\end{center}
\caption{The pair correlation function $g(r\!>\!1)$ of a
three-dimensional system of monodisperse spheres of radius 1,
illustrates the abundance of near contacts close to jamming
($\Delta \phi=10^{-8}$ here). From \cite{silbertPRE06} ---
Copyright by the American Physical Society. }\label{fig:gr}
\end{figure}

{\em Pair Correlation Function ---} In simulations of monodisperse
spheres in three dimensions, it was found that near jamming $g(r)$
diverges when $r \downarrow 1$ (for particles of radius 1):
\begin{equation}
g(r)\sim\frac{1}{\sqrt{r-1}}~.
\end{equation}
This expresses that at jamming a singularly large number of
particles are on the verge of making contact (Fig.~\ref{fig:gr})
\cite{silbertPRE06,silbert2002}. This divergence has also been
seen in pure hard sphere packings ~\cite{donevgr}. In addition to
this divergence, $g(r)$ exhibits a delta peak at $r=1$
corresponding to the $dN/2$ contacting pairs of particles.

In simulations of two-dimensional bidisperse systems, a similar
divergence can be observed, provided one studies $g(\xi)$, where
the rescaled interparticle distance $\xi$ is defined as
$r/(R_i+R_j)$, and where $R_i$ and $R_j$ are the radii of the
undeformed particles in contact  \cite{wouterzoranagr}.

\begin{figure}
\includegraphics[width=10.6cm,viewport=20 20 530 220,clip]{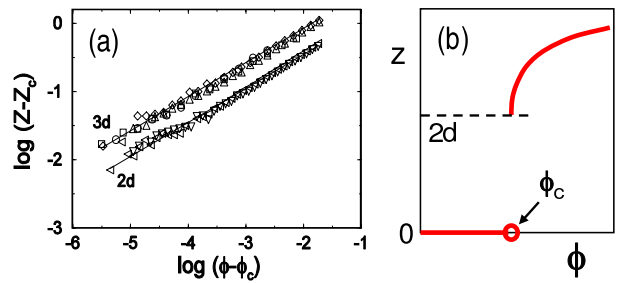}
\caption{(a) Excess contact number $z-z_c$ as function of excess
density $\phi-\phi_c$. Upper curves: represent monodisperse and
bidisperse packings of 512 soft spheres in three dimensions with
various interaction potentials, while lower curves correspond to
bidisperse packings of 1024 soft discs in two dimensions. The
straight lines have slope 0.5. From \cite{epitome} --- Copyright
by the American Physical Society. (b) Schematic contact number as
function of density, illustrating the mixed nature of the jamming
transition for frictionless soft spheres.}\label{fig:z}
\end{figure}

\subsection{Relating Contact Numbers and Packing Densities away
from J}\label{sec:awayJ}

Below jamming, there are no load bearing contacts and the contact
number is zero, while at point J, the contact number attains the
value $2d$. How does the contact number grow for systems at finite
pressure? Assuming that {\em(i)} compression of packings near
point J leads to essentially affine deformations, and that
{\em(ii)} $g(r)$ is regular for $r>1$, $z$ would be expected to
grow linearly with $\phi$: compression by 1\% would then bring
particles that are separated by less than 1\% of their diameter in
contact, etc. But we have seen above that $g(r)$ is not regular,
and we will show below that deformations are very far from affine
near jamming --- so how does $z$ grow with $\phi$?

Many authors have found that the contact number grows with the
square root of the excess density $ \Delta \phi:=\phi-\phi_c$
\cite{epitome,bolton,makseprl1999,durianbubble} (see
Fig.~\ref{fig:z}). O'Hern {\em et al.}~have studied this scaling
in detail, and find that the excess contact number $ \Delta z :=
z-z_{c}$ scales as $\Delta z \sim  (\Delta \phi)^{0.50\pm 0.03}$,
where $z_c$, the critical contact number, is within error bars
equal to the isostatic value $2d$ \cite{epitome}. Note that this
result is independent of dimension, interaction potential or
polydispersity (see Fig.~\ref{fig:z}a). Hence, the crucial scaling
law is
\begin{equation}
\Delta z = z_0 \sqrt{\Delta \phi}~,
\end{equation}
where the precise value of the prefactor $z_0$ depends on
dimension, and possibly weakly on the degree of polydispersity,
and is similar to $3.5\pm0.3$ in two dimensions and $7.9\pm0.5$ in
three dimensions \cite{epitome}.

The variation of the contact number near J can therefore be
perceived to be of mixed first/second order character: below
jamming $z=0$, at J the contact number $z$ jumps discontinuously
from zero to $2d$, and for jammed systems the contact number
exhibits non-trivial power law scaling as a function of increasing
density (Fig.~\ref{fig:durian} and \ref{fig:z}).

We will see below that many other scaling relations (for elastic
moduli, for the density of state and for  characteristic scales)
are intimately related to the scaling of $z$, and the contact
number scaling can be seen as the central non-trivial scaling in
this system. (In frictional and non spherical packings, similar
scalings for $z$ are found.)

A subtle point is that the clean scaling laws for $\Delta z$ vs
$\Delta \phi$ are only obtained if one excludes the rattlers when
counting contacts, but includes them for the packing fraction
\cite{epitome}. Moreover, for individual packings the scatter in
contact numbers at given pressure is quite substantial
--- see for example Fig.~9 from \cite{silkepreprint} --- and smooth curves
such as shown in Fig.~\ref{fig:z}a can only obtained by averaging
over many packings. Finally, the density $\phi$ is usually defined
by dividing the volume of the undeformed particles by the box
size, and packing fractions larger than 1 are perfectly
reasonable. Hence, in comparison to packing fractions defined by
dividing the volume of the {\em deformed} particles by the box
size, $\phi$ is larger because the overlap is essentially counted
double. Even though none of these subtleties should play a role
for the asymptotic scaling close to jamming in large enough
systems, they are crucial when comparing to experiments and also
for numerical simulations.

\subsubsection{Connections between contact number scaling, $g(r)$
and marginal stability}\label{subsub:con}

The scaling of $\Delta z$ can be related to the divergence of the
radial distribution function as follows \cite{silbert05}. Imagine
compressing the packing, starting from the critical state at point
J, and increasing the typical particle overlap from zero to
$\delta$. If one assumes that this compression is essentially
affine, then it is reasonable to expect that such compression
closes all gaps between particles that are smaller than $\delta$.
Hence
\begin{equation}
\Delta z \sim \int_1^{1+\delta} d \xi \frac{1}{\sqrt{\xi-1}} \sim
\sqrt{\delta }~.
\end{equation}

Wyart approaches the square root scaling of $\Delta z$ from a
different angle, by first showing that the scaling $\Delta
z\sim\sqrt\delta$ is consistent with the system staying marginally
stable at all densities, and then arguing that the divergence in
$g(r)$ is a necessary consequence of that~\cite{wyart05}. Both his
arguments require assumptions which are not self-evident, though
\cite{silkepreprint}.

\begin{figure}[t]
\includegraphics[width=6.0cm,viewport=15 10 230
210,clip]{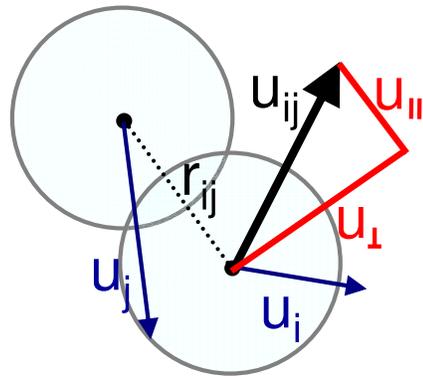} \caption{Definition of relative displacement
$u_{ij}$, $\uparl$ and $\uperp$. }\label{fig:defuparluperp}
\end{figure}

\subsection{Linear Response and Dynamical
Matrix}\label{sec:linresp}

A major consequence of isostaticity at point J is  that packings
of soft frictionless spheres exhibit increasingly anomalous
behavior as the jamming transition is approached. That anomalies
occur near jamming is ultimately a consequence of the fact that
the mechanical response of an isostatic system cannot be described
by elasticity --- isostatic systems are essentially different from
ordinary elastic systems  \cite{tchachenko,tcha}.

In principle these anomalies can be studied at the jamming point,
however, much insight can be gained by exploring the mechanical
properties as a function of distance to the isostatic point. Below
we review a number of such non-trivial behaviors and scaling laws
that arise near point J. We will focus on the response to weak
quasistatic perturbations, and on the vibrational eigenfrequencies
and eigenmodes of weakly jammed systems. Both are governed by the
dynamical matrix of the jammed packing under consideration.

For linear deformations, the changes in elastic energy can be
expressed in the relative displacement $\uij$ of neighboring
particles $i$ and $j$. It is convenient to decompose $\uij$ in
components parallel $(\uparl)$ and perpendicular $(\uperp)$ to
$\rij$, where $\rij$ connects the centers of particles $i$ and $j$
(Fig.~\ref{fig:defuparluperp}). In these terms the change in
energy takes a simple form~\cite{alexander,wyart05,wouterlinlang},
\begin{equation}
\label{dynmatdef} \Delta E=\frac12 \sum_{i,j} {k_{ij}} \left(~
u^2_{||,ij} -\frac{f_{ij}}{k_{ij} ~r_{ij}}u_{\perp,ij}^2\right)
 ~,
\end{equation}
where $f_{ij}$ and $k_{ij}$ denote the contact forces and
stiffnesses. For power law interactions of the form given in
Eq.~(\ref{eqs:interact}), we can rewrite this as
\cite{ellenbroek2006}:
\begin{equation}\label{DEPL}
\Delta E=\frac12 \sum_{i,j} k_{ij} \left(~ u^2_{||,ij}
-\frac{\delta_{ij}}{\alpha-1} u_{\perp,ij}^2\right)
 ~.
\end{equation}

The dynamical matrix $\dm_{ij,\alpha\beta}$ is obtained by
rewriting eq.~(\ref{dynmatdef}) in terms of the independent
variables, $u_{i,n}$, as
\begin{equation}
\label{dynmatdef2} \Delta E=\frac12 \dm_{ij,n m}~u_{i,n}~u_{j,m}
~.
\end{equation}
Here  $\dm$ is a $dN\times dN$ matrix with $N$ the number of
particles, indices $n,m$ label the coordinate axes, and the
summation convention is used.

The dynamical matrix contains all information on the elastic
properties of the system. By diagonalizing the dynamical matrix
one can probe the vibrational properties of systems near jamming
~\cite{epitome,tanguy,wyart05,silbert05} (see section
\ref{secdos}). The dynamical matrix also governs the elastic
response of the system to external forces $f^\mathrm{ext}$ (see
sections \ref{secscale}-\ref{secnonaf}
)~\cite{leonforte,ellenbroek2006}:
\begin{equation}
\label{responseeq} \dm_{ij,nm}~u_{j,m}=f^\mathrm{ext}_{i,n}~.
\end{equation}

\subsubsection{Density of States}\label{secdos}

\begin{figure}
\includegraphics[width=6.0cm,viewport=90 60 400
300,clip]{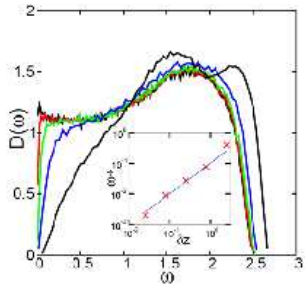} \caption{ Density of vibrational states
$D(\omega)$ for 1024 spheres interacting with repulsive harmonic
potentials. Distance to jamming $\Delta \phi$ equals $0.1$
(black), $10^{-2}$ (blue), $10^{-3}$ (green), $10^{-4}$ (red) and
$10^{-8}$ (black). The inset shows that the characteristic
frequency $\omega^*$, defined as where $D(\omega)$ is half of the
plateau value, scales linearly with $\Delta z$. The line has slope
1. Adapted from ~\cite{silbert05,wyart05} --- Copyright by the
Institute of Physics }\label{fig:dos}
\end{figure}

Studies of the vibrational modes, and the associated density of
(vibrational) states (DOS) have played a key role in identifying
anomalous behavior near point J. Low frequency vibrations in
ordinary crystalline or amorphous matter are long-wavelength plane
waves. Counting the number of these, one finds that the density of
vibrational states $D(\omega)$ is expected to scale as $D(\omega)
\sim \omega^{d-1}$ for low frequencies --- this is called Debye
behavior. Jammed packings of frictionless spheres do show
Debye-like behavior far away from jamming, but as the point J is
approached, both the structure of the modes and the density of
states exhibit surprising features
\cite{epitome,wyart05,silbert05,silbert09}.

The most striking features of the density of states are
illustrated in Fig.~\ref{fig:dos}. First, far above jamming, the
DOS for small frequencies is regular (black curve). Second,
approaching point J, the density of vibrational states DOS at low
frequencies is strongly enhanced. (In analogy to what is observed
in glasses, this is sometimes referred to as the boson peak, since
the ratio of the observed DOS and the Debye prediction exhibits a
peak at low $\omega$). More precisely, the DOS becomes essentially
constant up to some low-frequency crossover scale at
$\omega=\omega^*$, below which the continuum scaling $\sim
w^{d-1}$ is recovered. Third, the characteristic frequency
$\omega^*$ vanishes at point J as $\omega^* \sim \Delta z$.

The density of states thus convincingly shows that, close to the
isostatic point / jamming point, the material is anomalous in that
it exhibits an excess of low frequency modes, and that at point J,
the material does not appear to exhibit any ordinary
Debye/continuum behavior as here the DOS becomes flat. Jamming of
frictionless spheres thus describes truly new physics.

\begin{figure*}
\includegraphics[width=18.cm,viewport=1 1 600
140,clip]{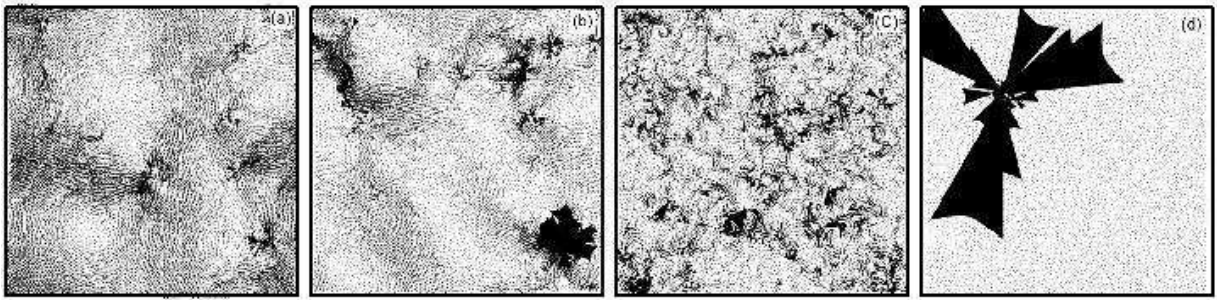} \caption{Representative eigenmodes for a
two-dimensional system of $10^4$ particles interacting with
three-dimensional Hertzian interactions ($\alpha=5/2$, see
Eq.~\ref{eqs:interact}) at a pressure far away from jamming
$(z\approx5.09)$. For all modes, the length of the vectors
$\propto u_i$ are normalized such that $\sigma_i |u_i|^2$ is a
constant. (a) Continuum-like low frequency mode at $\omega \approx
0.030, {\cal P} \approx 0.79$, and $i_{\omega}=3$, where
$i_{\omega}$ counts the non trivial modes, ordered by frequency.
(b) Quasi-localized low frequency mode at $\omega \approx 0.040,
{\cal P} \approx 0.06$, and $i_{\omega}=7$. (c) Disordered,
``swirly'' mid frequency mode at $\omega \approx 0.39, {\cal P}
\approx 0.31$, and $i_{\omega}=1000$. (d) Localized high frequency
mode at $\omega \approx 4.00, {\cal P} \approx 0.0013$, and
$i_{\omega}=9970$. }\label{fig:modes}
\end{figure*}

{\em Normal Modes ---} The nature of the vibrational modes changes
strongly with frequency, and, to a lesser extend, with distance to
point J. Various order parameters can be used to characterize
these modes, such as the (inverse) participation ratio, level
repulsion and localization length \cite{silbert09,zeravcic08}. The
participation ratio for a given mode is defined as ${\cal P}=
(1/N)~  (\Sigma_i  |u_i|^2)^2 / \Sigma_i |u_i|^4 $, where $u_i$ is
the polarization vector of particle $i$ \cite{silbert09}. It
characterizes how evenly the particles participate in a certain
vibrational mode
--- extended modes have  ${\cal P}$ of order one,
while localized modes have smaller ${\cal P}$, with hypothetical
modes where only one particle participates reaching ${\cal
P}=1/N$.

Studies of such order parameters have not found very sharp changes
in the nature of the modes either with distance to jamming or with
eigenfrequency \cite{silbert09,zeravcic08,vitelliunpub}. It
appears to be more appropriate to think in terms of typical modes
and crossovers. Qualitatively, one can consider the DOS to consist
of roughly three bands: a low frequency band where $D(\omega) \sim
\omega^{d-1}$, a middle frequency band where $D(\omega)$ is
approximately flat, and a high frequency band where $D(\omega)$
decreases with $\omega$ \cite{silbert09}.

Representative examples of modes in these three bands are shown in
Fig.~\ref{fig:modes}. The modes in the low frequency band come in
two flavors: plane wave like with ${\cal P} \sim 1$, and quasi
localized with small ${\cal P}$ \cite{zeravcic08,vitelliunpub}.
The modes in the large frequency band are essentially localized
with small ${\cal P}$. The vast majority of the modes are in the
mid frequency band (especially close to jamming), and are extended
but not simple plain waves
--- typically the eigenvectors have a swirly appearance.

The localization length $\xi$ of these modes has been estimated to
be large, so that many modes have $\xi$ comparable or larger than
the system size. Consistent with this, the modes in the low and
mid frequency range are mostly extended, $\xi>L$, and exhibit
level repulsion (i.e., the level spacing statistics $P(\Delta
\omega)$ follows the so-called Wigner surmise of random matrix
theory), while the high frequency modes are localized $(\xi<L)$
and exhibit Poissonian level statistics \cite{zeravcic08}.

When point J is approached, the main change is that the low
frequency, ``Debye'' range shrinks, and that both the number of
plane waves and of quasi-localized resonances diminishes
\cite{zeravcic08,silbert09,vitelliunpub}.

\subsubsection{Characteristic Length and Time Scales}\label{secscale}

The vanishing of the characteristic frequency $\omega^*$ at point
J suggests to search for a diverging length scale. Below we give
an analytical estimate for this length scale and discuss indirect
and direct observations of this length scale in simulations.

{\em Estimate of $l^*$ ---} As pointed out by Wyart {\em et al.}~
\cite{wyart05}, if we cut a circular blob of radius $\ell$ from a
rigid material, it should remain rigid. The rigidity (given by the
shear modulus) of jammed materials is proportional to $\Delta z$.
The circular blob has of the order $\ell^d\Delta z$ excess
contacts. By cutting it out, one breaks the contacts at the
perimeter, of which there are of order $z\ell^{d-1}$. If the
number of broken contacts at the edge is larger than the number of
excess contacts in the bulk, the resulting blob is not rigid, but
floppy: it can be deformed without energy cost (in lowest order).
The smallest blob one can cut out without it being floppy is
obtained when these numbers are equal, which implies that it has
radius $\ell^*\sim {z}/{\Delta z}$. Close to the jamming
transition, $z$ is essentially constant, and so one obtains as
scaling relation that \cite{wyart05}
\begin{equation}
\label{deflstar} \ell^*\sim \frac{1}{\Delta z}~.
\end{equation}

{\em Observation of $l^*$ in Vibration Modes ---} Using the speed
of sound one can translate the crossover frequency $\omega^*$ into
a wavelength, which scales as $\lambda_\mathrm{T}\sim
1/\sqrt{\Delta z}$ for transverse (shear) waves and as
$\lambda_\mathrm{L}\sim 1/\Delta z$ for longitudinal
(compressional) waves --- the difference in scaling is due to the
difference in scaling of shear and bulk moduli (see section
\ref{secmodul} below). By examining the spatial variation of the
eigenmode corresponding to the frequency $\omega^*$,
$\lambda_\mathrm{T}$ has been observed  by Silbert \emph{et
al.}~\cite{silbert05}. Notice, however, that the scaling of
$\lambda_\mathrm{T}$ is different from the scaling of $l^*$ --- it
is $\lambda_\mathrm{L}$ that coincides with the length scale
$\ell^*$ derived above.

\begin{figure}
\includegraphics[width=8.6cm,viewport=40 40 520
450,clip]{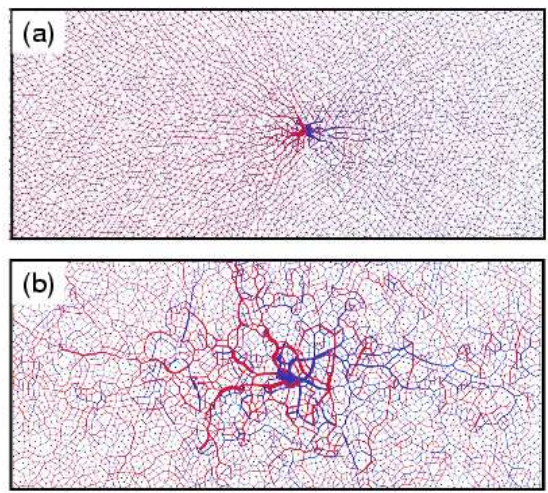} \caption{Divergence of a characteristic
length scale  near jamming as observed in the fluctuations of the
changes of contact forces of a system of $10^4$ Hertzian discs.
Blue (red) bonds correspond to increased (decreased) force in
response to pushing a single particle in the center of the packing
to the right. In panel (a), the system is far from jamming and
$z=5.55$, while in panel (b), the system is close to jamming and
$z = 4.05$ (adapted from \cite{wouterlinlang}).
 }\label{fig:lwouter}
\end{figure}

{\em Observation of $l^*$ in Point Response ---} The signature of
the length scale $\ell^*$ can be observed directly in the point
force response networks : Close to point J, i.e. for small $\Delta
z$,  the scale up to which the response looks disordered becomes
large (see Fig.~\ref{fig:lwouter})
\cite{ellenbroek2006,wouterlinlang}. By studying the radial decay
of fluctuations in the response to an inflation of a single
central particle (which is more symmetric than that of point
forcing as shown in Fig.~\ref{fig:lwouter}) as a function of
distance to jamming, one obtains a crossover length $l^*$ which,
as the theoretically derived length scale, varies as $l^* \approx
6/\Delta z$ \cite{wouterlinlang}.

{\em Characteristic Length and Validity of Elasticity ---} An
important issue, which has in particular been studied extensively
in the context of granular media, is whether elasticity can
describe a systems response to, for example, point forcing
\cite{tcha,goldenberg}. Extensive observations of the linear
response, connected to the direct observation of $l^*$, suggest
that there is a simple answer, and that the distance to the
isostatic limit is crucial \cite{ellenbroek2006,wouterlinlang}:
Below a length scale $l^*$ the response is dominated by
fluctuations, and the deformation field can be seen as a distorted
floppy mode, while at larger length scales the systems response
crosses over to elasticity. This is for a single realization ---
it can also be shown that, even close to jamming, the ensemble
averaged response of a weakly jammed system is consistent with
elasticity, provided the correct values of the elastic moduli are
chosen --- these moduli are consistent with the globally defined
ones \cite{wouterlinlang}.

\subsubsection{Scaling of Shear and Bulk Moduli}\label{secmodul}

\begin{figure}[t]
\includegraphics[width=8.4cm,viewport=30 45 460 260,clip]{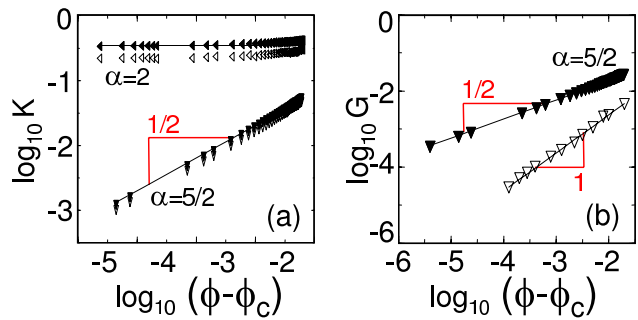}
    \caption{Bulk (K) and shear (G) modulus as function of
    distance to jamming for two-dimensional bidisperse systems, with
    interaction potential $V \sim \delta^{\alpha}$ (see Eqs.~\ref{eqs:interact}).
    The closed symbols denote moduli calculated by forcing the
    particles to move affinely, and the open symbols correspond to
    the moduli calculated after the system has relaxed. Slopes as
    indicated (adapted from \cite{epitome} --- Copyright by the American Physical Society).}
    \label{fig:GK}
\end{figure}

The scaling of the shear modulus, $G$, and bulk modulus, $K$,
plays a central role in connecting the non-affine, disordered
nature of the response to the anomalous elastic properties of
systems near jamming. To understand why disorder is so crucial for
the global, mechanical response of collection of particles that
act through short range interactions, consider the local motion of
a packing of spherical, soft frictionless spheres under global
forcing. The global stresses can be obtained from the relative
positions $\vec{r}_{ij}$ and contact forces $\vec{f}_{ij}$ of
pairs of contacting particles $i$ and $j$ via the Irving-Kirkwood
equation:
\begin{equation}
\Sigma_{\alpha \beta} = \frac{1}{2V}  ~\Sigma_{ij} f_{ij,\alpha}
r_{ij,\beta}~,
\end{equation}
where $\sigma_{ab}$ is the stress tensor, $\alpha$ and $\beta$
label coordinates, and $V$ is the volume.

Once we know the local motion of the particles in response to an
externally applied deformation, we can calculate the contact
forces from the force law and obtain thus the stress in response
to deformation. Let us first estimate the scaling of the moduli
from the affine prediction where one assumes that the typical
particle overlap $\delta$ is proportional to $\Delta \phi$ and
that all bonds contribute similarly to the increase in elastic
energy when the packing is deformed. For a deformation strain
$\varepsilon$ we can estimate the corresponding increase in energy
from Eq.~(\ref{DEPL}) as $\Delta E \sim \Sigma k \varepsilon^2$.
Therefore, under affine deformations, the corresponding elastic
modulus is of order $k$
--- in other words, the elastic moduli simply follow from the
typical stiffnesses of the contacts.

Consider now deforming a {\em disordered} jammed packing. All
particles feel a local disordered environment, and deformations
will not be affine (Fig.~\ref{fig:affine}). The point is that
these non-affine motions become increasingly strong near the
jamming transition, and qualitatively change the scaling behavior
of, e.g., the shear modulus of foams and granular media
\cite{epitome,bolton,makseprl1999,alexander,wouterepl}.

A particularly enlightening manner to illustrate the role of
non-affine deformations is to initially force the particle
displacements to be affine, and then let them relax. In general,
the system can lower its elastic energy by additional non-affine
motions. Calculating the elastic energies of enforced affine
deformations and of the subsequent relaxed packings of soft
frictionless spheres, O'Hern and co-workers found that the
non-affine relaxation lowers both the shear and bulk modulus, but
crucially changes the scaling of the shear modulus with distance
to jamming \cite{epitome} --- see Fig.~\ref{fig:GK}.

\begin{figure*}[t]
\includegraphics[width=18.6cm,viewport=15 10 830
260,clip]{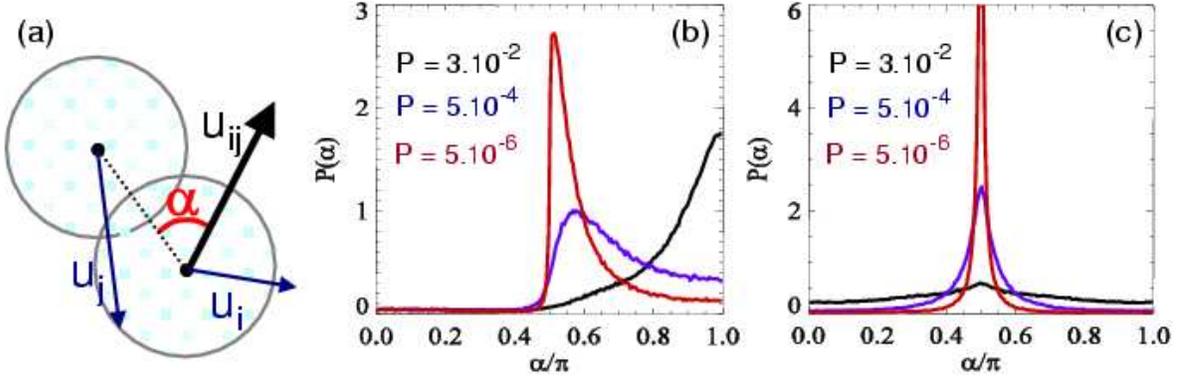} \caption{(a) Illustration of definition of
displacement angle $\alpha$. (b-c) Probability distributions
$P(\alpha)$ for compression (b) and shear (c), for Hertzian
particles in two dimensions. The three pressures indicated
correspond to $z\approx 6.0$,   $z\approx 4.5$ and $z\approx 4.1$
respectively (adapted from \cite{ellenbroek2006} --- Copyright by
the American Physical Society).}\label{fig:alpha}
\end{figure*}

In general, one finds that for power law interactions
(Eq.~\ref{eqs:interact}), the pressure scales as $\Delta
\phi^{\alpha-1}$ and the contact stifness $k$ and bulk modulus $K$
scale as $\Delta \phi^{\alpha-2}$ \cite{epitome,ellenbroek2006,
wouterepl}. The surprise is that the shear modulus $G$ gets
progressively smaller than the bulk modulus near point J, and $G$
scales differently from $K$ with distance to jamming: $ G \sim
\Delta \phi^{\alpha-3/2}$ (See Fig.~\ref{fig:GK})
\cite{epitome,makseprl1999,ellenbroek2006, wouterepl}. The
relations between the scaling of $G$, $K$ and $k$ can be rewritten
as
\begin{equation} \label{gdzk}
G \sim \Delta z  K \sim \Delta z ~\!k~.
\end{equation}
It is worth noting that many soft matter systems (pastes,
emulsions) have shear moduli which are much smaller than
compressional moduli
--- from an application point of view, this is a crucial property.

Putting all this together, we conclude that the affine assumption
gives the correct prediction for the bulk modulus (since $k \sim
\delta^{\alpha-2} \sim \Delta \phi^{\alpha-2}$), but fails for the
shear modulus. This failure is due to the strongly non-affine
nature of shear deformations: deviations from affine deformations
set the elastic constants
\cite{alexander,epitome,makseprl1999,ellenbroek2006, wouterepl}.
As we will see below, the correspondence between the bulk modulus
and the affine prediction is fortuitous, since the response
becomes singularly non-affine close to point J, for both
compressive and shear deformations (section \ref{emtrp}).

\subsubsection{Non-Affine Character of Deformations}
\label{secnonaf}

Approaching the jamming transition, the spatial structure of the
mechanical response  becomes less and less similar to continuum
elasticity, but instead increasingly reflects the details of the
underlying disordered packing and becomes increasingly non-affine
\cite{ellenbroek2006} --- see Fig.~\ref{fig:affine}a. Here we will
discuss this in the light of Eq.~(\ref{DEPL}), which expresses the
changes in energy as function of the local deformations $\uparl$
and $u_{\perp}$ : $\Delta E=\frac12 \sum_{i,j} k_{ij} \left(~
u^2_{||,ij} -\frac{\delta_{ij}}{\alpha-1} u_{\perp,ij}^2\right)$.

To capture the degree of non-affinity of the response, Ellenbroek
and co-workers have introduced the displacement angle
$\alpha_{ij}$ \footnote{not to be confused by the power law index
of the interaction potential}. Here $\alpha_{ij}$ denotes the
angle between $\vc{u}_{ij}$ and $\vc{r}_{ij}$, or,
\begin{equation}
\label{defalpha}
\tan\alpha_{ij}=\frac{u_{\perp,ij}}{u_{\parallel,ij}}~.
\end{equation}

The probability distribution $P(\alpha)$ can probe the degree of
non-affinity by comparison with the expected $P(\alpha)$ for
affine deformations. Affine compression corresponds to a uniform
shrinking of the bond vectors, i.e. $u_{\perp,ij}=0$ while
$u_{\parallel,ij}=-\varepsilon r_{ij}<0$: the corresponding
$P(\alpha)$ exhibits a delta peak at $\alpha=\pi$. The effect of
an affine shear on a bond vector depends on its orientation, and
for isotropic random packings, $P(\alpha)$ is flat.

Numerical determination of $P(\alpha)$ show that systems far away
from the jamming point exhibit a $P(\alpha)$ similar to the affine
prediction, but that as point J is approached, $P(\alpha)$ becomes
increasingly peaked around $\alpha=\pi/2$
(Fig.~(\ref{fig:alpha})b-c). This is reminiscent of the
$P(\alpha)$ of floppy deformations, where the bond length does not
change and $P(\alpha)$ exhibits a $\delta$-peak at $\pi/2$. Hence
deformations near jamming become strongly non-affine, and, at
least locally, resemble those of floppy modes.

{\em Non-affinity of Floppy Modes and Elastic Response ---} Wyart
and co-workers have given variational arguments for deriving
bounds on the energies and local deformations of soft (low energy)
modes starting from purely floppy (zero energy)
modes~\cite{wyart05,wyart08}. They construct trial soft modes that
are basically floppy modes, obtained by cutting bonds around a
patch of size $\ell^*$, and then modulating these trial modes with
a sine function of wavelength $\ell^*$ to make the displacements
vanish at the locations of the cut bonds
~\cite{wyart05,ellenbroek2006}. In particular, for the local
deformations, they find~\cite{wyart08}
\begin{equation}
\frac{\uparl}{\uperp}\sim \frac{1}{\ell^*} \rightarrow
\frac{\uparl}{\uperp} \sim \Delta z ~, \label{scale_prediction1}
\end{equation}
where symbols without indices ${ij}$ refer to typical or average
values of the respective quantities.

The question is whether the linear response follows this
prediction for the soft modes. The width $w$ of the peak in
$P(\alpha)$ is, close to the jamming transition, roughly
$\uparl/\uperp$ because $|\alpha_{ij}-\pi/2|\approx
u_{\parallel,ij}/u_{\perp,ij}$ if $u_{\parallel,ij}\ll
u_{\perp,ij}$. It turns out that the scaling behavior
(\ref{scale_prediction1}) is consistent with the width $w$ of the
peak of $P(\alpha)$ for shear deformations, but not for
compression. There the peak of $P(\alpha)$ does not grow as much,
and a substantial shoulder for large $\alpha$ remains even close
to jamming: the tendency for particles to move towards each other
remains much more prominent under compression.

{\em Scaling of $\uparl$ and $\uperp$ ---} The scaling of the
distributions of $\uparl$ and $\uperp$ has also been probed. The
key observation is that in Eq.~(\ref{DEPL}) the terms $\sim
u_{||}$ and $u_{\perp}$ have opposite signs. What is the relative
contribution of these terms, and can we ignore the latter?
Surprisingly, even though $\delta \ll 1$,
Eq.~(\ref{scale_prediction1}) predicts that the two terms are of
equal magnitude in soft modes, and so for linear response one
needs to be cautious.

It has become clear that the balance of the terms is never so
precise as to qualitatively change the magnitude of the energy
changes: $\Delta E$ and $\frac12 \sum_{i,j} k_{ij} (~
u^2_{||,ij})$ scale similarly \cite{wouterlinlang,wouterepl}.
Hence, the typical values of $\uparl$ under a deformation are
directly connected to the corresponding elastic modulus: For
compression $ \uparl$ is essentially independent of the distance
to jamming $(\uparl \sim \epsilon)$, while for shear, $\uparl \sim
\epsilon ~ \Delta \phi ^{1/4}$, where $\epsilon$ is  the magnitude
of the strain \cite{wouterlinlang,wouterepl}.

The scaling for $\uperp$, the amount by which particles in contact
slide past each other, is more subtle. Numerically, one observes
that for shear deformations, $\uperp \sim \epsilon ~ \Delta \phi
^{-1/4}$. The two terms $\sim u_{||}$ and $\sim u_{\perp}$ become
comparable here, and the amount of sideways sliding under a shear
deformation diverges near jamming
\cite{ellenbroek2006,wouterepl,wouterlinlang}. For compression
there is no simple scaling. Combining the observed scaling for
$\uparl$ with Eq.~(\ref{scale_prediction1}), one might have
expected that $\uperp \sim \epsilon ~ \Delta \phi ^{-1/2}$.
However, the data suggests a weaker divergence, close to $\Delta
\phi ^{-0.3}$. Hence, consistent with the absence of simple
scaling of the peak of $P(\alpha)$ for compression, the two terms
$\propto u_{||}$ and $\propto u_{\perp}$ do not balance for
compression. Nevertheless, both under shear and compression, the
sliding, sideways motion of contacting particles dominates and
diverges near jamming.

\subsubsection{Effective Medium Theory, Rigidity Percolation,
Random Networks and Jammed Systems} \label{emtrp}

In 1984, Feng and Sen  showed that elastic percolation is not
equivalent to scalar percolation, but forms a new universality
class \cite{Feng}. In the simplest realization of rigidity
percolation, bonds of a ordered spring network are randomly
removed and the elastic response is probed. For such systems, both
bulk and shear modulus go to zero at the elastic percolation
threshold \footnote{To translate the data for c11 and c44 as
function of p shown in fig 1, note that G = C44, and K = c11 -
c44. All go to zero linearly in $p-p_c$.}, and at this threshold
the contact number reaches the isostatic value $2d$ \cite{Feng2}.
Later it was shown that rigidity percolation is singular on
ordered lattices \cite{GenRP}, but similar results are expected to
hold on irregular lattices.

While it has been suggested that jamming of frictionless spheres
corresponds to the onset of rigidity percolation [59], there are
significant differences, for example that the contact number
varies smoothly through the rigidity percolation threshold but
jumps at the jamming transition \cite{epitome}. Nevertheless, it
is instructive to compare the response of random spring networks
of given contact number to those of jammed packings --- note that
the linear response of jammed packings of particles with one-sided
harmonic interactions is exactly equivalent to that of networks of
appropriately loaded harmonic springs, with the nodes of the
network given by the particle centers and the geometry and forces
of the spring network determined by the force network of the
packing.

\begin{figure}
\hfill\includegraphics[width=8.3cm,viewport=10 0 570
350,clip]{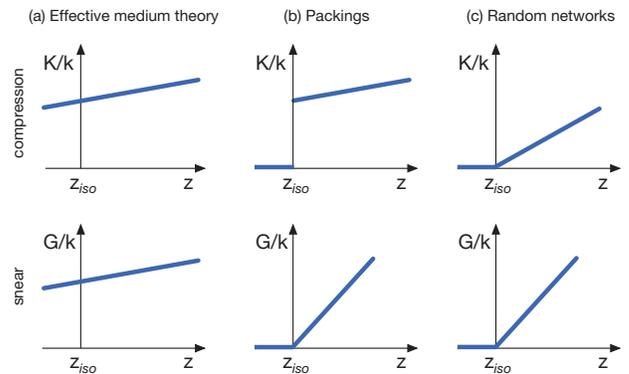} \caption{Schematic comparison of the
variation of shear ($G$) and bulk ($K$) elastic moduli as function
of distance to jamming. (a) In effective medium theory, all
elastic moduli are simply of the order of the local spring
constant $k$, and moreover, the theory does not account for
whether the packing is rigid or not. (b) In jammed packings of
harmonic particles, the bulk modulus $K$ remains constant down to
the jamming transition, where it vanishes discontinuously, whereas
the shear modulus $G$ vanishes linearly in $\Delta z$. (c) In
random networks of elastic springs, both elastic moduli vanish
linearly with $\Delta z$. (From \cite{wouterepl} --- Copyright by
the Institute of Physics). }\label{fig:emtjampack}
\end{figure}

In Fig.~(\ref{fig:emtjampack}), a schematic comparison of the
variation of the elastic moduli with contact number in effective
medium theory, for jammed packings and for random networks is
shown. This illustrates that EMT predicts that the elastic moduli
vary smoothly through the isostatic point and that the moduli are
of order of the local spring constant $k$. This is because
effective medium theory is essentially ``blind'' to local packing
considerations and isostaticity. Thus, besides failing to capture
the vanishing of $G$ near jamming, its prediction for the bulk
modulus fails spectacularly as well: it predicts finite rigidity
below isostaticity. Clearly random networks also fail to describe
jammed systems, as for random networks both shear and bulk modulus
vanish when $z$ approaches $z_{\rm iso}$ --- from the perspective
of random networks, it is the bulk modulus of jammed systems that
behaves anomalously.

By comparing the displacement angle distributions $P( \alpha)$ of
jammed systems and random networks under both shear and
compression, Ellenbroek {\em et al.}~conclude that two cases can
be distinguished \cite{wouterepl}. In the ``generic'' case, all
geometrical characterizations exhibit simple scaling and the
elastic moduli scale as $\Delta z$ --- this describes shear and
bulk deformations of randomly cut networks, as well as shear
deformations  of jammed packings. Jammed packings under
compression form the ``exceptional'' case: the fact that the
compression modulus remains of order $k$ near jamming is reflected
in the fact that various characteristics of the local
displacements do \emph{not} exhibit pure scaling.

\begin{figure}
\hfill\includegraphics[width=8.6cm,viewport=0 00 750
450,clip]{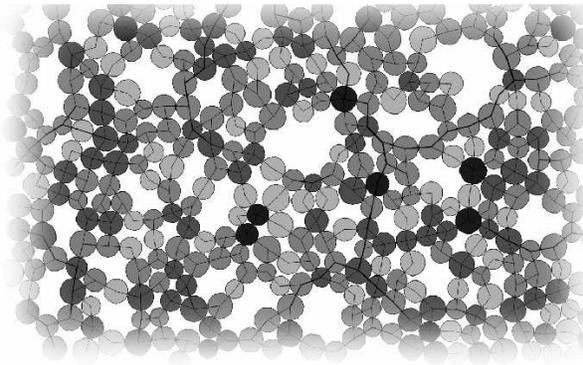} \caption{Part of a packing of frictional
discs in two dimensions for low pressure, zero gravity and
friction coefficient $\mu=10$. For this packing, $z \approx 3.06$
and $\phi \approx 0.77$ (this density includes rattlers, which are
not shown in this image and occur in the ``holes''). Lines
indicate the strength of the normal forces --- note the large
number of near contacts (pairs of particles appearing to touch but
not connected by a force line). Disc color indicates local contact
number, clearly identifying the large fraction of particles with
two contacts only --- these do not arise in frictionless systems.
}\label{fig:fricsnap}
\end{figure}

\subsection{Conclusion}

For packings of soft frictionless spheres and in the limit of
large systems, contact number, packing density, particle
deformation and (for given force law) pressure are all directly
linked and at point J the system becomes isostatic. The jamming
transition for frictionless spheres exhibits a number of
non-trivial scaling behaviors, all intimately linked to the
non-trivial square-root scaling of the excess contact number with
distance to the isostatic jamming point. We have stressed the
viewpoint that geometry and mechanics are intimately linked for
these systems, and that near point J, local non-affinity and
global anomalous mechanical scaling go hand in hand.

\section{Jamming of Frictional Spheres}\label{sec:fric}

Here we discuss the rich phenomenology of jamming of frictional
soft spheres. The crucial difference with the frictionless case is
that both the packing density $\phi_c$ and contact number $z_c$ at
jamming  are not unique: both depend on the friction coefficient
and on the history of the packing, and are lower than for
frictionless spheres
\cite{makse2000,silbert2002,kasahara,unger2005,zhang2005,shundjak2007,somfai2007}
--- see figure~\ref{fig:fricsnap}.

\begin{figure}
\hfill\includegraphics[width=10.6cm,viewport=20 20 380
145,clip]{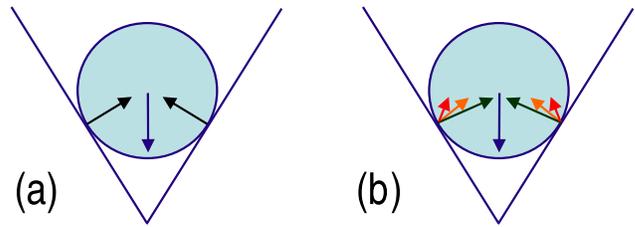} \caption{Frictionless (a) and frictional (b)
disc in a groove \cite{ballinagroove}. (a) In the frictionless
case, the system is isostatic, and the contact forces (black)
balancing the gravitational force (blue) are unique. (b) In the
frictional case, the system is hyperstatic: contact forces in hard
frictional systems are, in general, under-determined. In this
example, there are four force degrees of freedom (two normal and
two frictional forces), and only three balance equations (total
force in $x$ and $y$ direction and torque balance). This leads to
a family of solutions (three examples indicated in red, orange and
green) that balance the gravitational force (blue). Which of these
is realized depends on the history of the system.}
\label{fig:ballgroove}
\end{figure}

Jamming and isostaticity no longer go hand in hand for frictional
spheres. The contact number at jamming, $z_c$, can range from
$d+1$ to $2d$, where $d+1$ is the isostatic value $z_{\rm
iso}^{\mu}$ for frictional spheres (see section \ref{iso_fric} and
Appendix \ref{sec:count}). It appears that $z_c$ approaches
$z_{\rm iso}^{\mu}$ only in the limit of $\mu \rightarrow \infty$
and very slow equilibration of the packings
\cite{zhang2005,shundjak2007,somfai2007} --- see section
\ref{sec:geniso}. In all other cases, the number of contacts at
jamming is larger than the minimal number needed for force balance
and rigidity, and frictional packings of soft spheres at jamming
(or, equivalently, frictional rigid spheres), are hyperstatic:
$z_c > z_{\rm iso}^{\mu}$. Hyperstaticity implies that for
packings of rigid, frictional spheres, the contact forces are not
uniquely determined by the packing geometry, as was the case for
the isostatic packings of rigid, frictionless spheres
\cite{moukarzel,tchachenko}. An explicit example of this so-called
indeterminacy of frictional forces is shown in
Fig.~\ref{fig:ballgroove} \cite{ballinagroove}.

What does the deviation of the critical contact number from the
isostatic value imply for the scaling of quantities such as $G, K$
and $\omega^*$? We will show that these scale with {\em distance
to the frictional isostatic point}, $z-z_{\rm iso}^{\mu}$. Thus,
when the jamming transition is approached, bulk quantities in
general do not exhibit scaling with distance to the jamming point,
since, at jamming, $z$ approaches $z_c \ge z_{\rm iso}^{\mu}$
\cite{makse2000,silbert2002,zhang2005,shundjak2007,somfai2007,silbertprivate2008}.
Scaling with distance to jamming can only occur when $z_c = z_{\rm
iso}^{\mu}$. Hence, jamming is not critical for frictional
systems: power law scaling of bulk quantities with distance to
jamming is the exception, not the rule.

The jamming scenario for frictional soft spheres is detailed
below. We briefly discuss the frictional contact laws in section
\ref{sec:friclaw}. In section \ref{ssec:fric} we discuss the
properties of frictional sphere packings at the jamming threshold,
or equivalently, packings of undeformable frictional spheres. We
focus on the variation of the range of contact numbers and
densities as function of $\mu$ in sections
\ref{iso_fric}-\ref{sec:musc}. Finally in section \ref{sec:geniso}
we introduce the concept of generalized isostaticity, which is
relevant for frictional packings that have fully mobilized
contacts. Section \ref{sec:mup} concerns frictional packings at
finite pressures and we discuss the (breakdown) of scaling with
distance to jamming.

\subsubsection{Frictional Contact Laws}\label{sec:friclaw}

Friction is taken into account by extending the contact force
model to account both for normal forces ${F_{\rm n}}$ and
tangential forces $F_{\rm t}$. In the simple Coulomb picture of
friction, contacts do not slide as long as the ratio of tangential
and normal forces remains smaller than or equal to the friction
coefficient $\mu$: $|F_{\rm t}|/F_{\rm n} \le \mu$, which
introduces a very sharp nonlinearity in the contact laws. Typical
values for $\mu$ relevant in experiments range from 0.1 to 1,
which is where properties of frictional packs vary strongly with
$\mu$.

Frictional forces do not only depend on the relative position of
the contacting particles, but also on their history
\cite{johnson,ballinagroove,kasahara,maksepre2004,somfai2007,silbertprivate2008}.
This is encoded in the widely used Hertz-Mindlin model for
frictional three-dimensional spheres, which takes the normal force
$F_{\rm n} \sim \delta^{3/2}$ with $\delta $ the overlap between
particles, while the tangential force {\em increment} ${\rm d}
F_{\rm t} \sim \delta^{1/2} {\rm d}t$ where ${\rm d}t$ is the
relative tangential displacement change, provided $F_{\rm t}\leq
\mu F_{\rm n}$ \cite{johnson,maksepre2004,somfai2007}. Studies of
friction can also be performed for other contact laws, most
notably, the linear model for which $F_{\rm n} \sim \delta$, so
that the stiffness of the contacts in the normal and tangential
direction are independent of the normal force and do not vary with
distance to jamming \cite{silbert2002}.

\subsection{Frictional Packings at Zero Pressure}\label{ssec:fric}

\subsubsection{Contact Number}\label{iso_fric} How can the
counting arguments for the contact number at zero pressure be
extended to the frictional case? On the one hand, the requirement
that contacting spheres precisely touch is the same as for the
frictionless case, and gives $zN/2$ constraints on the $dN$
particle coordinates, leading to $z \leq 2d$. On the other hand,
for {\em frictional} packings, the constraint counting for the
$zdN/2$ contact force components constrained by $dN$ force and
$d(d-1)N/2$ torque balance equations (see Appendix
\ref{sec:count}) gives $z \geq d+1$, where the isostatic value
$z^\mu_{\rm iso}=d+1$. Combining these two bounds, frictional
spheres can attain a range of contact numbers: $d+1 \leq z_c \leq
2d$ (see Appendix \ref{sec:count}).

\begin{figure}
\hfill\includegraphics[width=10.6cm,viewport=50 50 550
240,clip]{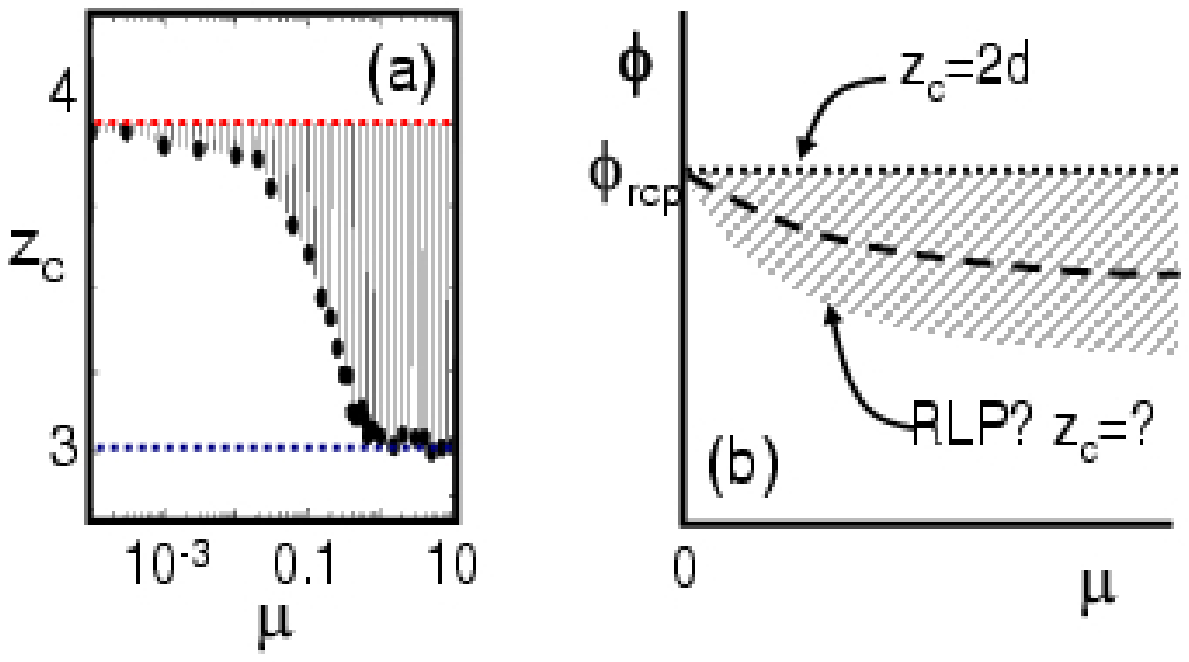} \caption{(a) Example of the variation of the
zero pressure contact number $z_c$ in two-dimensional rigid discs
as function of $\mu$, smoothly interpolating between the isostatic
limits $2d$ (red) for zero friction and $d+1$ (blue) for
frictional contacts. The arched area indicate combinations of
contact numbers and $\mu$ that, while they are not reached in
these numerics, are perfectly possible --- see text (adapted from
\cite{unger2005} --- Copyright by the American Physical Society).
(b) State diagram for frictional spheres. While the Random Close
Packed, isostatic packings obtained for zero friction are
compatible with all values of $\mu$, a range of packings with
lower densities and contact numbers open up when $\mu >0$. For a
given preparation protocol, there might be a well-defined density
(dashed curve). Whether there is a well defined lowest packing
fraction for given $\mu$, which would define Random Loose Packing,
is an open question, and the question what the contact number of
such states would be is open as well ( adapted from
\cite{0510506v1.pdf}).}\label{fig:friczphi}
\end{figure}

It is important to stress that neither bound is sensitive to the
value of $\mu$. What mechanism (if any) selects the contact number
$z_{c}$ of a frictional packing at jamming? The first additional
ingredient to consider is the Coulomb criterion that for all
contact forces $|F_t|/F_n \leq \mu$. So, while constraint counting
allows force configurations that satisfy force and torque balance
for $z$ arbitrarily close to $z^{\mu}_{\rm iso}$, such
configurations are not guaranteed to be compatible with the
Coulomb criterion, and in particular for small $\mu$ they
generally will not be. This is consistent with the intuition that
a small increase of $\mu$ away from zero is not expected to make
$z_c$ jump from $2d$ to $d+1$. In section \ref{sec:geniso} we will
discuss an additional bound on $z$ as function of $\mu$.

Simulations show that in practice $z_c$ is a decreasing function
of $\mu$, approaching $2d$ at small $\mu$ and approaching $d+1$
for large friction coefficient
\cite{silbert2002,unger2005,zhang2005,shundjak2007,somfai2007}
(Fig.~\ref{fig:friczphi}a). However, $z_c(\mu)$ cannot be a
sharply defined curve unless additional information about the
preparation history is given: From the non-sliding condition
$|F_t|/F_n \le \mu$ it follows that a packing which is stable for
a certain value of ${\mu}$ remains so for all larger values of
${\mu}$ --- increasing the friction coefficient only expands the
range of allowed force configurations (and does not change any of
the contact forces). Hence, a numerically obtained curve
$z_c(\mu)$ at best is a bound for the allowed combinations of
$z_c$ and $\mu$ (see Fig.~\ref{fig:friczphi}a). History is a
second additional ingredient to consider \cite{kasahara}, although
it is remarkable that several different equilibration algorithms
appear to give very similar estimates for $z_c(\mu)$
\cite{makse2000,silbert2002,kasahara,unger2005,zhang2005,shundjak2007,somfai2007}.

\subsubsection{Density} The existence of a wide range of
statistically different frictional packings is also reflected in
packing densities, which experimentally are more easily observed
than the contact number. It is well known that packings of
spherical hard particles under gravity (in other words, frictional
spheres close to jamming) can be compacted over a range of
densities \cite{compaction}. Different packing densities of these
systems do not correspond to deformations of the particles, but to
changes in the organization of the particles. Hence, at jamming,
the range of packing densities does not go to zero for frictional
particles.

The relation between density and friction coefficient can be
summarized in a simple state diagram (Fig.~\ref{fig:friczphi}b),
which stresses that random close packing (RCP) is independent of
$\mu$, while the random loose packing (RLP) density depends
strongly on $\mu$, thus connecting random close packing, random
loose packing and value of the friction coefficient
\cite{bernal,onoda,
zhang2005,shundjak2007,maksenature2008,ciamarra2008,silbertprivate2008}.
This diagram further suggests that the packing density at point J
may also be seen as random loose packing of frictionless spheres
(since for $\mu=0$, one expects RCP and RLP to coincide)
--- it is the loosest possible packings, rather than the densest
possible ones, that arise near jamming. It should be noted that
the definition of RLP is even more contentious than RCP, and the
debate is wide open
\cite{maksenature2008,ciamarra2008,silbertprivate2008,ciamarra2008}.

\subsubsection{Scaling with $\mu$}\label{sec:musc}

One may now also  wonder how the contact number and packing
density at jamming scale with $\mu$. Qualitative evidence for
scaling was found by Silbert {\em et al.}~in numerical studies of
frictional packings (Fig.~2 and 3 from \cite{silbert2002}). By
focussing explicitly on a single preparation protocol, such as
slow equilibration, this becomes a well posed question --- leading
to the concept of generalized isostaticity, defined below. Data
for generalized isostatic packings suggests that both contact
number and density exhibit power law scaling with $\mu$ for small
friction, while for large friction, excess contact number and
density (defined with respect to the infinite friction limit) are
also related by scaling, although clearly more work is needed to
establish these scalings firmly
\cite{shundjak2007,silbertprivate2008}.

\begin{figure}
\includegraphics[width=8.4cm,viewport=230 10 550
215,clip]{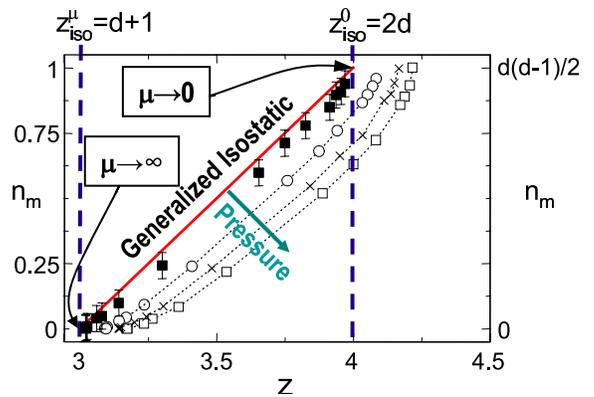} \caption{Generalized isostaticity plot,
comparing the fraction of fully mobilized contacts per particle,
$n_m$, to the contact number, $z$. Data points (open symbols) are
for two-dimensional systems and for $\mu$ ranging from $0.001$ to
$1000$ at finite $P$. The black squares are the corresponding
$n_m$ and $z$ extrapolated to $P=0$. The left and bottom axes
refer to the numerical values for contact number and number of
fully mobilized contacts per particle, $n_m$, for this specific
two-dimensional example, while right and top axes give the
corresponding general expressions for higher dimensions. The red
line denotes the generalized isostaticity line where the number of
fully mobilized contacts is maximized: $ n_m = d(z -d-1)/2$. The
area to the right of this line refers to generalized hyperstatic
packings, while the area to the left of the red line is forbidden
(adapted from \cite{shundjak2007} --- Copyright by the American
Physical Society). }\label{fig:geniso}
\end{figure}

\subsubsection{Generalized Isostaticity}\label{sec:geniso}

Here we will discuss the role of the frictional forces in some
more detail, and in particular focus on frictional packings for
which a large number of contacts are fully mobilized, meaning that
the frictional forces are maximal: $|F_t|/F_n = \mu$). These
packings arise in numerical studies when packings are equilibrated
slowly for a wide range of values of $\mu$.

The mobilization, $m$, of a contact is defined as the ratio
$|F_t|/(\mu F_n)$, and  ranges from zero to one (fully mobilized).
Earlier numerical data suggested that $m$ generally stays away
from 1, and that in the limit of large $\mu$, the distribution of
the mobilization $P(m)$ becomes independent of $\mu$
\cite{silbert2002,kasahara}. Later it became clear that $P(m)$ can
depend strongly on the preparation history \cite{zhang2005}.
Futhermore, frictional two-dimensional packings which are very
slowly equilibrated yield packings for which a substantial amount
of the contact forces are fully mobilized, meaning that $|F_t|/F_n
= \mu$ \cite{shundjak2007,silbert2002,bouchaud}. One imagines that
during equilibration, many contacts slowly slide, and when the
packing jams many contacts are still close to failure --- such
packings are marginal with respect to lowering $\mu$.

For packings with fully mobilized contacts, the counting arguments
need to be augmented, since at fully mobilized contacts, the
frictional and normal forces are no longer independent
\cite{shundjak2007}. Defining the number of fully mobilized
contacts per particle as $n_m$, the constraints for the $zdN/2$
force degrees of freedom then are: $dN$ force balance equations,
$d(d-1)N/2$ torque balance equations, and $n_m N$ constraints for
the fully mobilized contacts. This yields the following relation
between $z$, $z^\mu_{\rm iso}=d+1$ and $n_m$\footnote{the
corresponding equation in \cite{shundjak2007} is only correct for
$d=2$.}:
\begin{equation}\label{genisobound}
z-z^\mu_{\rm iso}\ge 2 n_m/d ~.
\end{equation}

Surprisingly, for sufficiently slowly equilibrated packings and
for all values of $\mu$, the values for $n_m$ and $z$ tend to
satisfy this bound when $P$ is lowered to zero
(Fig.~\ref{fig:geniso}). Such packings which maximize their number
of fully mobilized contacts have been referred to as ``generalized
isostatic'' packings \cite{bouchaud,shundjak2007}. These should be
widely occurring, since most preparation algorithm tend to
equilibrate slowly
\cite{makse2000,silbert2002,kasahara,unger2005,zhang2005,shundjak2007,somfai2007}.

For fully mobilized packings, the amount of fully mobilized
contacts vanishes in the limit of infinite friction (see
Fig.~\ref{fig:geniso}), consistent with the observation that there
$z \approx d+1$. The number of fully mobilized contacts is maximal
for vanishingly small friction (which we refer to as $ \mu=0^+$),
where, by continuity, $z\approx 2d$, and $n_m \approx d(d-1)/2$.
Taking into account that each contact is shared by two particles,
the fraction of fully mobilized contacts is $(d-1)/2$ --- hence in
two dimensions, 50\% of all contacts are fully mobilized, in three
dimensions, 100\% of the contacts would be fully mobilized for
$\mu=0^+$, and in higher dimensions one cannot reach generalized
isostaticity for $\mu=0^+$.

By itself, the inequality (\ref{genisobound}) is not a stricter
bound on $z$ than the ordinary condition $z\ge d+1$, since
$n_m(\mu)$ is unknown. However, if we could determine $n_m(\mu)$,
we would immediately obtain the bound $z=d+1+2 ~n_m(\mu)/d$. It
is, at present, an open question how $n_m(\mu)$ can be estimated
or obtained numerically other than through direct numerical
simulations.

\begin{figure}
\hfill\includegraphics[width=8.8cm,viewport=20 20 455
505,clip]{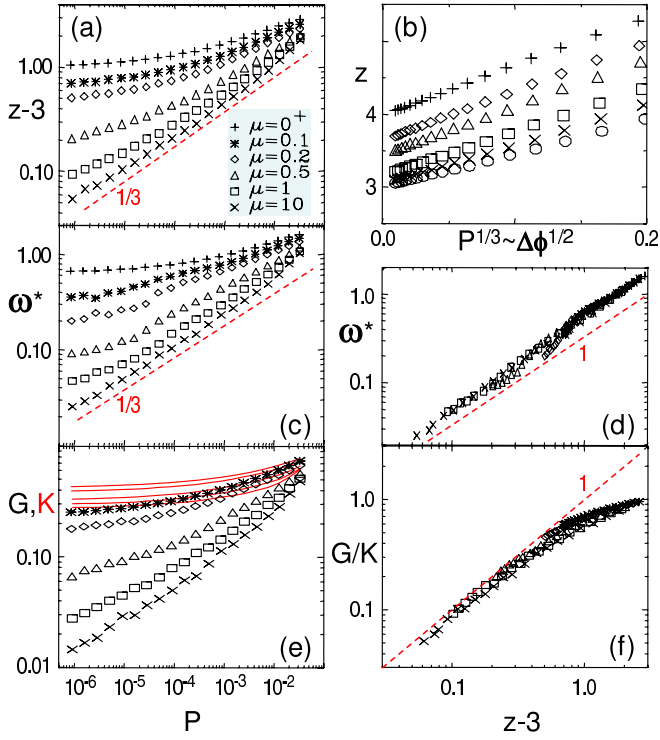} \caption{Scaling of contact number, $w^*$
and elastic moduli for frictional discs, interacting through
three-dimensional Hertzian-Mindlin forces. (a) The zero pressure
contact number, $z_J$ does not reach the isostatic limit $(z=3)$
unless $\mu$ is very large. (b) The excess coordination number
$z-z_J$ scales linearly with $P^{1/3}\sim \sqrt{\Delta \phi}$.
(c-d) The characteristic frequency of the DOS, $\omega^*$, scales
similarly to $z-3$ \cite{p16}. (e) The bulk modulus $K$ (red
curves) approaches a plateau for small $P$, while $G$ appears to
scale as $z-3$ \cite{p13}. (f) As in frictionless spheres, the
ratio $G/K$ scales with distance to the isostatic point, now given
by $z-z_{\rm iso}^{\mu}=z-3$ (adapted from \cite{somfai2007} ---
Copyright by the American Physical Society
).}\label{fig:fricscale}
\end{figure}

\subsection{Frictional Packings at Finite Pressure}\label{sec:mup}

Once a mechanically stable frictional packing has been created,
its linear mechanical response is given by the dynamical matrix.
For Hertz-Mindlin type interactions, each contact can be thought
of as being given by two springs (one parallel to the contact
vector $r_{ij}$, one perpendicular to the contact vector), the
spring constants of which are set by the normal force and the
Poisson ratio \cite{Hertzian}.

Various authors have found that, for essentially all values of
$\mu$, the excess  contact number $z-z_c$ grows as a square root
with the excess density \cite{makse2000,somfai2007,
silbertprivate2008}
--- for Hertzian contacts, this is equivalent to stating that
$z-z_c \sim P^{1/3}$. However, $z_c$ differs from the frictional
isostatic value $d+1$, so that $z-z_{\rm iso}^{\mu}$ does {\em
not} scale with pressure (see Fig.~\ref{fig:fricscale}a,b). Note
that the slope in Fig.~\ref{fig:fricscale}b, which represents the
prefactor $z_0$ in a scaling law of the form $z-z_c =z_0
\sqrt{\phi-\phi_c}$ does not appear to vary strongly with $\mu$.
As is the case for frictionless particles, it is essential to
remove rattlers for the count of the contact number, but include
them for the estimate of the density to obtain the square root
scaling  of $z-z_c$ over an appreciable range \cite{shundjak2007}.
This square root scaling is intriguing and, as far as we are
aware, without explanation.

The deviations of $z_c$ from the isostatic value imply that
packings near the (un)jamming transition do not approach isostatic
packings, and consistent with this, there is, in general,  no
scaling of the mechanical properties as function of distance to
jamming.

The mechanical properties do, however, scale with the distance to
the isostatic point, as measured by the contact number. First,
calculations of the characteristic frequency $\omega^*$ from the
density of vibrational states for two-dimensional frictional
packings show that the variation of $\omega^*$  with $\mu$ and
distance to jamming (as measured by the pressure $P$) is very
similar to that of $z$ (Fig.~\ref{fig:fricscale}c). In fact, when
this data is replotted as function of $z-z_{\rm iso}^{\mu}=z-3$
one finds a linear relation between $\omega^*$ and $z-z_{\rm
iso}^{\mu}$ (Fig.~\ref{fig:fricscale}d). Second, the ratio of the
shear and bulk modulus exhibits the same phenomenology: $G/K$
scales linearly with $z-z_{\rm iso}^{\mu}=z-3$
(\ref{fig:fricscale}e-f) \cite{somfai2007,makseEPL2008}. These
findings suggest that, in general, scaling is governed by the
distance to isostaticity, rather than the distance to jamming.

The contact number and geometry of the packings change smoothly
with $\mu$ \cite{somfai2007,silbertprivate2008}, while the
mechanical behavior exhibits a discontinuous jump from $\mu=0$ to
$\mu=0^+$. This is caused by the fact that when friction is
included, the nature of the dynamical matrix changes completely,
because the tangential contact stiffnesses jump from zero to a
finite value. When the tangential stiffness is varied smoothly
from $\mu=0$ to finite friction, the mechanical properties vary
smoothly also \cite{silkepreprint2}.

Finally a word of caution regarding the notion of generalized
isostaticity and the role of fully mobilized contacts  for scaling
away from jamming. In the calculations presented above fully
mobilized contacts are treated as ordinary elastic contacts.
Strictly speaking, such marginal contacts cause a breakdown of
linear response. One may argue that tiny perturbations would
simply let the fully mobilized contacts relax to
almost-fully-mobilized, after which linear response would no
longer be problematic. Taking the opposite view, Henkes {\em et
al.}~have recently shown that if the dynamical matrix is
calculated under the assumption that fully mobilized contacts can
slide freely, the characteristic frequency $\omega^*$ scales and
vanishes with $P$ for all values of $\mu$ --- provided one
considers systems that approach generalized isostaticity
\cite{silkepreprint2}.

\subsection{Conclusion}\label{mudis}

Jamming of frictional grains can be seen as a two-step process.
The first step is the selection of a contact number, $z$, given
the friction coefficient, pressure and procedure. In the second
step, in which the mechanical properties of the packing are
determined, everything scales with $z-z_{\rm iso}^{\mu}$. The
crucial difference with frictionless spheres is that the contact
number $z_c$ at the $P=0$ jamming point in general does not
coincide with $z_{\rm iso}^{\mu}$. Most quantities are governed by
the contact number and scale with distance to isostaticity, while
the contact number itself scales with distance to jamming.

\section{Jamming of Non-Spherical Particles}\label{sec:ellipsoid}

New phenomena occur in packings of non-spherical particles, and
here we briefly discuss the jamming scenario for frictionless
ellipsoids.

\begin{figure}
\includegraphics[width=6.3cm,viewport=80 30 420 500,clip]{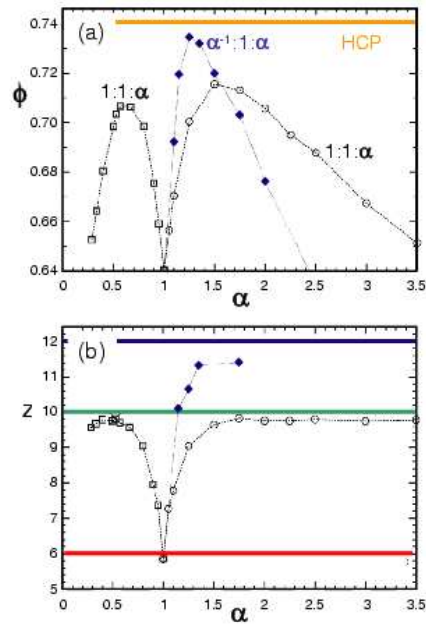}
\caption{(a) Packing fraction  $\phi$ of spheroids (open symbols)
and general ellipsoids (blue symbols) as function of the
asphericity $\alpha$. The density shows a cusp at $\alpha=1$
(sphere limit). The orange line indicates the HCP packing density
$\approx 0.74$, which is almost reached by random packings of some
ellipsoids. (b) The contact number, $z$, for the same spheroids
and ellipsoids as shown in panel (a) also shows a cusp at
$\alpha=1$. The red, green and blue lines at $z=6$, $z=10$ and
$z=12$ indicate the isostatic contact numbers for spheres,
spheroids and ellipsoids (adapted from \cite{donev2004science}
--- Copyright by The American Association for the Advancement of Science). }\label{ellipsoid1}
\end{figure}

First, configurations for  hard (or zero pressure) frictionless
ellipsoids pack more densely and have larger contact numbers than
frictionless spheres
\cite{donev2004science,donev2007,coreyellipses,zz2009}. As we
discuss in section~\ref{ellipsoidal.pack}, both the increase in
density and in contact number away from the sphere limit are
continuous but not smooth --- plots of $\phi$ and $z$ as function
of the ellipticity show a cusp at the sphere limit
(Fig~\ref{ellipsoid1}).

Second, the counting arguments for general ellipsoids suggest that
at jamming, ellipsoids attain $z=z_{\rm iso}= d(d+1)$. However,
weakly aspherical ellipsoids actually attain a contact number
arbitrarily close to the sphere limit $z=2d$. As a consequence,
(weakly) ellipsoidal packings are strongly hypostatic
(underconstrained) near jamming. This leads to questions about the
relation between contact number, rigidity and floppy modes
(section \ref{sec:countel}).

Third, the question arises whether quantities such as $z$ and
$\omega^*$ exhibit scaling, either as function of the pressure, as
function of the asphericity or as function of distance to either
the spherical or the ellipsoidal isostatic point --- the partial
answers to these questions, based on recent studies of the density
of states \cite{coreyellipses,zz2009} will be addressed in section
~\ref{sec:presel}.

\subsection{Packings of Spherocylinders, Spheroids and
Ellipsoids}\label{ellipsoidal.pack}

{\em Spherocylinders ---} Early indications for surprisingly dense
packings of non-spherical particles come from studies of
spherocylinders, particles consisting of a cylinder of length  $a$
and diameter $1$, which on both ends are capped by a half sphere.
For zero $a$, these are spheres, while the large $a$ limit is
relevant for the loose packings of thin (colloidal) rods
\cite{philipse1996}. Williams {\em et al.}~studied the packing
fraction and contact numbers of such spherocylinders numerically,
and found that both the packing fraction $\phi$ and contact number
$z$ increase when $a$ is increased, reach a maximum for $a \sim
0.4$, and then decrease \cite{williams2003}. The density peaks at
a value of 0.695, substantially larger than the typical values for
random close packing of spheres $\sim 0.64$, while for large $a
>10$ the density decays as $\phi \sim 1/a$, consistent with
arguments given before \cite{philipse1996}.

The contact number in these simulations was found to start out at
$z \approx 5.8 $ for $a=0$, and increased until it reached $z
\approx 9$ for $a\approx 0.4$. The initial value is close to the
isostatic number for spheres (6), while the peak value is similar
to the isostatic number for rods (10) \cite{philipseprivcom}.

{\em Spheroids and Ellipsoids ---} In seminal work, Donev {\em et
al.}~explored the packing properties of hard spheroids and
ellipsoids \cite{donev2004science}. As shown in
Fig~\ref{ellipsoid1}a, the density of spheroids (axis:
$1:1:\alpha$) exhibits a cusp-like local minimum for the pure
spherical case $\alpha=1$, and reaches two local maxima: oblate
(disc-like) spheroids  at $\alpha \approx 0.6 $ pack at a density
$\phi \approx 0.70$ and prolate (cigar-shaped) ellipsoids at
$\alpha=1.5$ pack even denser at $\phi \approx 0.715$. Note that
the spheroid packing density only drops below the random close
packing value for spheres for very strongly oblate ($\alpha
\lesssim 0.25$) or prolate ($\alpha \gtrsim 4$) particles.

Even larger packing densities can be obtained for triaxial
ellipsoids, and for the case that the axes are given as
$1/\alpha:1:\alpha$, the maximum packing density peaks at 0.735
for $ \alpha \approx 1.25$,  \cite{donev2004science,donev2007}.
This density is surprisingly close to the density $\approx 0.74$
obtained for fcc and hcp packings, which are the densest possible
packings for spheres --- but those are crystals, whereas the
ellipsoidal packings do not show any appreciable orientational
ordering. Finally, crystals of ellipsoids can be packed even
denser, with the highest density currently known, 0.7707, is
obtained in non lattice periodic packings of spheroids with either
$\alpha \ge \sqrt{3}$ or $\alpha \le 1/\sqrt{3}$
\cite{donev2004prl}.

The contact number grows monotonically with asphericity, from a
value $\approx 2d$ for the spherical case to values close to the
corresponding higher isostatic number for ellipsoids: The contact
number for the spheroids measured for strongly oblate or prolate
appears to level off at values around 9.8 (the corresponding
isostatic number is 10), and for ellipsoids one reaches 11.4 (the
corresponding isostatic number is 12) \cite{donev2004science}.
(The contact numbers in the disordered ellipsoidal systems are
difficult to obtain accurately from numerics, in particular for
hard particles
--- since, similar to hard spheres, one expects anomalously many
near contacts \cite{silbertPRE06,donevgr}). It is noteworthy that
the contact numbers reach these asymptotic values at the same
asphericities where the density is maximal. Recent work on
two-dimensional ellipses \cite{donev2007,coreyellipses} and three
dimensional spheroids \cite{zz2009} confirm these trends in
contact numbers.

\subsection{Counting arguments, Floppy modes and Rigidity of
Ellipsoids}\label{sec:countel}

The counting arguments for general ellipsoids suggest that at
jamming, ellipsoids attain $z=z_{\rm iso}= d(d+1)$. However,
weakly aspherical ellipsoids actually attain a contact number
arbitrarily close to the sphere limit $z=2d$. Hence counting
arguments suggest that packings of weakly ellipsoidal particles
possess a large number of floppy modes. Are these packings stable?

As a first step in understanding such packings, it is helpful to
think about weakly aspherical ellipsoids that approach the sphere
limit. The number of floppy modes in such an underconstrained
system equals $(N/2)(z_{\rm iso}-z)$, which for the sphere limit
(where $z \rightarrow 2d$) equals $N (d(d-1)/2)$. What are these
floppy modes?

The key observation is that in the counting arguments for
ellipsoids, the rotational degrees of freedom are taken into
account while for spheres, where they correspond to trivial
rotations of the particles, these are ignored. When these
rotational degrees of freedom are also taken into account for
frictionless spheres, one obtains precisely $N (d(d-1)/2)$ trivial
floppy modes, corresponding to the trivial rotational degrees of
freedom of individual frictionless spheres
\cite{zz2009,donev2007}. These floppy modes do not affect the
rigidity of the packings, which suggests that, in general, absence
of floppy modes may be a sufficient but not a necessary condition
for rigidity \cite{donev2007}.

From the perspective of constraint counting of the contact forces,
something similar happens in the sphere limit: how do $dN$ force
degrees of freedom satisfy both $dN$ force balance equations and
also all the additional torque balance equations? The answer is
simple: for frictionless spheres, the torques exerted by each
contact force is zero, and so torque balance is trivially
satisfied.

The key question, however, is what happens to hypostatic packings
at {\em finite} asphericity and pressure. The full answers are not
known, but two recent studies on the density of vibrational states
for soft frictionless bidisperse two-dimensional ellipses
\cite{coreyellipses} and three-dimensional spheroids \cite{zz2009}
provide important ingredients that we will discuss below.

\subsection{Jamming of Ellipsoids}\label{sec:presel}

\begin{figure}
\hfill \includegraphics[width=9cm,viewport=10 0 530
150,clip]{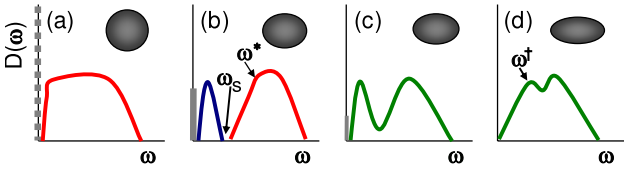} \caption{Schematic scenario for the density
of states for frictionless soft ellipsoidal particles, based on
\cite{coreyellipses,zz2009}. (a-d) Density of states as function
of distance to the spherical limit. The grey, blue, red and green
colors refer to floppy modes, rotational modes, translational
modes and hybridized modes respectively. (a) For frictionless
spheres, one usually only considers the translational band (red),
but when one takes the rotational degrees of freedom into account,
a large number of trivial floppy modes occur (dashed grey line).
(b) For contact numbers just above $z-z_{\rm iso}^{\rm sphere}$,
the density of states exhibits three bands, and the characteristic
frequencies $\omega_s$ and $\omega^*$ scale with asphericity and
$z-z_{\rm iso}^{\rm sphere}$ respectively --- see text. (c) For
contact numbers approaching $z-z_{\rm iso}^{\rm ellip}$, the
rotational and translational band merge. (d) For contact numbers
above $z-z_{\rm iso}^{\rm ellip}$, there are no floppy modes and
the characteristic frequency $\omega^{\dagger}$ scales with
$z-z_{\rm iso}^{\rm ellip}$}\label{figzz1}
\end{figure}

The main findings for the density of states of ellipsoidal
particles are shown in Fig.~\ref{figzz1}. Close to the sphere
limit, where the contact number is far below the relevant
ellipsoidal isostatic value, the density of states consists of
three bands: first, a number of zero frequency, floppy modes
corresponding to the degree of hypostaticity, second, a band of
rotational modes, and third, a band of translational modes,
corresponding to the translational modes present for the pure
sphere case. When, for increasing pressure and/or strong
ellipticity, the contact number starts to approach the ellipsoidal
isostatic value, the rotational and translational bands hybridize
and merge. Finally, when the contact number exceeds the
ellipsoidal isostatic value, the floppy modes have vanished and
the characteristic frequency of the remaining single band density
of states scales with distance to the ellipsoidal isostatic value.

\begin{figure}
\includegraphics[width=9cm,viewport=10 0 360 260,clip]{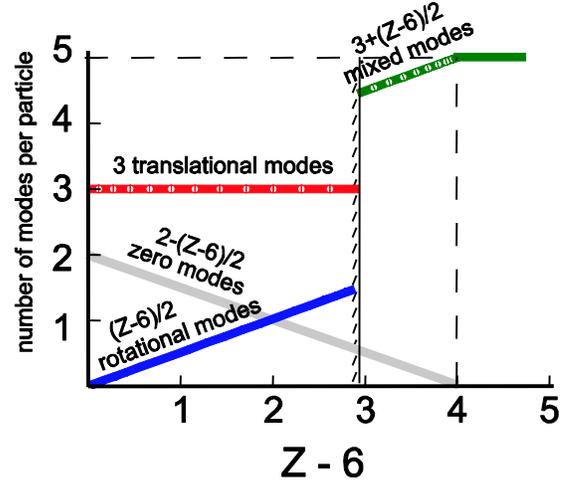}
\caption{Schematic representation of the number of modes per band
for the specific case of spheroids in three dimensions --- from
\cite{zz2009} --- Copyright by the Institute of
Physics}\label{figzz2}
 \end{figure}

The counting arguments provide a clear picture of the number of
modes per band, as shown in Fig.~\ref{figzz2}, where the variation
of these numbers with contact number is shown for the case of
spheroids in 3d.

First, the vibrational modes present for the spherical case are
only weakly perturbed by the inclusion of weak ellipticity, so
their number still equals $dN$. The particle motions of modes in
this band are essentially translational, and the characteristic
frequency of this band, $\omega^*$, still scales with $z-z_{\rm
iso}^{\rm sphere}$, {\em not} with $z-z_{\rm iso}^{\rm ellips}$.
Hence, this part of the density of states is smoothly perturbed
when going from the sphere to the weakly ellipsoidal case.

Second, for $z <z_{\rm iso}^{\rm ellip}$ the system is
underconstrained, and the crucial observation is that here there
are $(z - z_{\rm iso}^{\rm ellip})/2$ floppy modes. In the sphere
limit, these modes are the trivial local rotations, and away from
the sphere limit most of these modes survive and become
delocalized --- their precise nature is not fully understood yet.

Thirds, at finite pressures and/or finite asphericities,
$(z-z_{\rm iso}^{\rm sphere})/2$ modes emerge from the zero
frequency band and attain finite frequencies. This is the
rotational band: particle motions of modes in this band are
essentially rotational, and the vibration frequencies are below
those of the translational band. This allows the definition of a
characteristic maximal frequency of the rotational band
$\omega_s$, which is found to scale with the degree of asphericity
$|1-\alpha|$, but is essentially insensitive to the pressure.

Fourth, for large pressure and asphericity, the contact number
approaches the relevant ellipsoidal isostatic number, the
rotational and translational bands start to approach each other (
$\omega^*-\omega_s \ll 1$), the modes hybridize and these two
bands  eventually merge. In the regime where the contact number
exceeds the relevant ellipsoidal contact number, there are no more
floppy modes. The only band of vibrational modes then has a mixed
translational/rotational character, and its characteristic
frequency, $\omega^{\dagger}$, scales with distance to the
relevant ellipsoidal isostatic point: $\omega^{\dagger} \sim
z-z_{\rm iso}^{\rm ellip}$.

Finally, note that for weakly elliptical systems that are
hypostatic, the counting argument implies that the forces must be
non-generic --- one still has more equations of force and torque
balance than one has force degrees of freedom. In terms of the
elastic energy landscape, one imagines that near such systems
there must exhibit many directions in phase space where the second
derivative is zero (leading to quartic modes
\cite{coreyellipses}), but a deep understanding is missing.

\subsection{Conclusion}

Jamming of frictionless ellipsoidal particles is surprisingly
similar to that of frictionless spheres, despite the strongly
hypostatic nature of weakly aspherical packings. The crucial
observation is that frictionless spheres can also be seen as
strongly hypostatic near jamming, as they possess a large number
of trivial floppy modes. Most of these modes remain at zero
frequency for weakly ellipsoidal particles, even though their
spatial structure is no longer trivial, and these modes do not
appear to affect the rigidity of packings of frictionless
ellipses.

\section{Summary, Open Questions and Outlook}\label{sec:disc}

\begin{figure}
\includegraphics[width=8.6cm,viewport=100 0 650 520,clip]{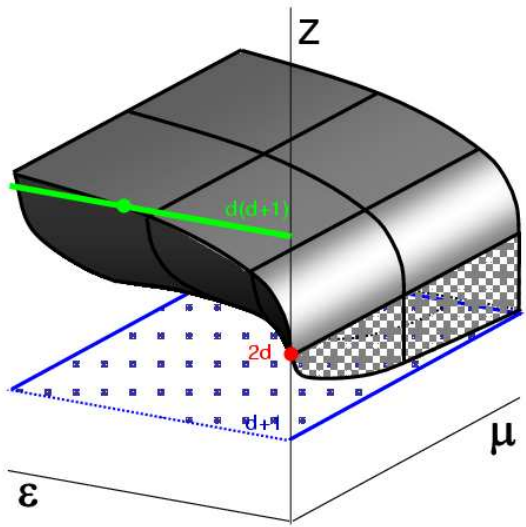}
\caption{Conjectured range of selected contact numbers at jamming,
i.e., at $P=0$, as function of the friction coefficient $\mu$ and
ellipticity $\epsilon$. The red dot indicates the isostatic limit
for frictionless spheres at ($\mu=0,\epsilon=0$), the green line
indicates the isostatic limit for frictionless ellipsoids
($\mu=0,\epsilon \neq 0$) and the blue plane indicates the
isostatic limit for frictional particles ($\mu \neq 0$). The
contact number is precisely selected in the frictionless plane,
%minor changes here
and for sufficiently large ellipticity the contact number crosses
the isostatic value (green dot). Once friction comes into play, a
range of contact numbers are allowed. For a given $\mu$ and
$\epsilon$, the upper bound is given by the selected contact
number for frictionless ellipsoids at $\mu=0$, while the lower
bound is given by the generalized isostaticity limit --- i.e., for
finite $\mu$, the maximal number of contacts is fully mobilized
here, and only for $\mu \rightarrow \infty$ does $z$ reach the
frictional isostatic value $z=d+1$.
 }\label{fig:z2}
 \end{figure}

The jamming scenario for disordered packings of soft, purely
repulsive particles at zero temperature and shear, as described
above, can be seen as a two step process. First, for a given
pressure, contact law and preparation protocol, a packing with a
certain contact number, $z$, is created. Second, the mechanical
characteristics such as elastic moduli and density of states
depend on the difference between the actual contact number and the
relevant isostatic value.

Depending on the particles' friction or shape, the contact number
may span a range of values --- see Fig.~\ref{fig:z2} for this
range for $P\rightarrow 0$. For frictionless particles it  appears
that the contact number at jamming is independent of the
preparation procedure, even for finite pressures. For frictional
particles, a range of contact numbers arises and the history
becomes crucial.

Jamming of frictionless soft spheres constitutes a special case,
since here the isostatic contact number (excluding the trivial
rotational degrees of freedom of the particles) is reached at the
jamming threshold. The counting for ellipsoidal particles takes
these rotational degrees into account, which leads to strongly
hypostatic packings near jamming --- however, the associated zero
modes do not appear to contribute to the mechanical properties of
the packings. Furthermore, the perturbation from spheres to weak
ellipsoids is smooth, when the trivial rotational modes for the
spheres is included.

Friction, however, acts differently. Given a certain preparation
procedure, the change in contact number is smooth with $\mu$.
However, the frictional interactions are such that at the level of
the dynamical matrix, the inclusion of arbitrary small friction
introduces a discontinuous change. For any value of the friction
the tangential stiffness takes on a finite value which leads to
contributions to the dynamical matrix of order one, contributions
which are absent in the frictionless case. Friction become a
smooth perturbation only when the tangential stiffness is varied
smoothly with $\mu$.

\subsection{Open Questions}

A crucial question is that of experimental relevance. Many recent
predictions of the theory should be observable in experiment, in
particular for frictionless systems such as foams and emulsions,
but very few have been observed so far. Frictional packings have
been explored theoretically far less than frictionless systems,
despite their obvious experimental relevance
\cite{behringerdz,jacob}. How many different order parameters does
one need to characterize the statistics of generic frictional
packings?

More work is needed to clarify the notion of random loose packing
\cite{onoda,maksenature2008,ciamarra2008}, and to unravel the role
of packing protocols. What is the underlying distribution of
possible contact numbers and densities for frictional spheres,
given a certain pressure and friction coefficient? Do RCP and RLP
correspond to sharp gradients in this distribution? Are the RCP
and RLP limits identical for frictionless packings? Random
packings of spheres are much looser then random packings of non
spherical particles --- can we understand why?

It is, in many cases, unknown how results obtained for
frictionless spheres extend to more complex systems. For example,
do a diverging length scale and a singularly non-affine response
arise when frictional spheres or ellipses approach their isostatic
limit(s)? What about the elastic moduli
\cite{somfai2007,makseEPL2008}? Similarly, what is the jamming
scenario for more general particles, such as frictional ellipses
and non convex particles that may share multiple contacts? What is
the scenario for more general interactions (attraction, long
range...)?

Given the central role of the square-root scaling of the contact
number with distance to jamming, it would be useful to probe the
connection to the square-root singularity of $g(r)$ --- the
argument outlined in section~\ref{subsub:con} assumes
displacements to be primarily affine, while near $J$, the
displacements are singularly non-affine and diverge. What may
happen is that the relative displacement of particles that are not
in contact are not strongly non-affine --- we don't know. For
frictional spheres it is not understood whether $z-z_c$ exhibits
true square root scaling with excess density, and whether $g(r)$
exhibits similar scaling behavior there.

Essentially all the work discussed above focuses on averaged
quantities and linear response. For finite systems, contact
numbers, moduli etc exhibit significant differences in different
realizations \cite{epitome,silkepreprint}. Can we understand these
fluctuations near jamming?  What is the nonlinear yielding
behavior of systems near jamming \cite{epitome}?

A whole host of new phenomena arise when jammed systems are put
under shear stress, and possibly are made to flow
\cite{olsson,langlois,head,hatano08}, or when systems of finite
temperature \cite{berthier,vestige} are considered. Can these
phenomena be connected in a meaningful manner to the zero shear,
zero temperature limit?

\subsection{Outlook}

Jamming is cool \cite{jamnote}, as it provides a framework to
approach the mechanics of disordered systems. The studies of the
simplest case of static soft frictionless spheres have
demonstrated that such systems exhibit rich spatial organization
and anomalous mechanical properties near the isostatic/jamming
limit. Important tasks for the coming years include  exploring the
relevance of these observations for experimental observations and
for systems with more complex interactions. New horizons are
emerging for systems at finite temperature and in particular for
flow near jamming --- as attested by the rich phenomenology of
flowing foams, suspensions and granular media.

%\end{multicols}%Mvh formatting

\newpage
\appendix

\section{Counting Arguments for the Contact Number}\label{sec:count}

By constraint counting one can establish bounds on the contact
number \cite{alexander}. First, one may require that floppy modes,
deformations that in lowest non trivial order do not cost energy,
are absent. This yields a lower bound on the contact number.
Packings that violate this second constraint are called
hypostatic, packings that marginally fulfill this constraint are
isostatic, and packings that fulfill this constraint are called
hyperstatic.

Note that the same lower bound on the contact number is obtained
by requiring that all contact forces balance. As we will see, this
is because the number of independent degrees of freedom
necessarily to describe changes in the energy at a contact equals
the number of force degrees of freedom per contact. Therefore, the
requirement that floppy modes are absent is equivalent to the
requirement that the contact forces balance, and often the
counting argument that yields the lower bound on $z$ is phrased in
terms of the contact forces.

Secondly, for packings at jamming, one arrives at a second
constraint, which follows from the requirement that the particles
are undeformed at jamming. This yields an upper bound on the
contact number. Violations of this second condition are possible
for special (non-generic) packings, such as perfect crystals.

As we will see,  for frictionless particles the first and second
bounds coincide. This does not necessarily imply that the
corresponding contact numbers are {\em realized} at jamming:
numerically it is found that frictionless spheres are indeed
isostatic at jamming \cite{epitome}, while weakly aspherical
frictionless ellipsoids are strongly hypostatic
\cite{donev2007,coreyellipses,zeravcic08}. For frictional
particles the two bounds never coincide, and numerically it is
found that frictional particles are almost always hyperstatic at
jamming.

Below we present the counting arguments in detail, for packings of
$N$ soft particles in $d$ dimensions which interact through
contact forces, and for which the contact number $z$, is defined
as the average number of contacts per particle. Note that the
total number of contacts equals $N z/2$ --- each contact is shared
by two particles. As we will find below, to perform these counting
arguments we need to know the number of force components per
contact, or equivalently, the number of independent degrees of
freedom necessarily to describe changes in the energy at a contact
($\tilde{f}$), the geometrical number of degrees freedom per
particle ($\tilde{x}$) and the number of force balance equations
per particle ($\tilde{b}$).

\begin{table*}[htb]
\begin{tabular} {l|l|l|l|l|l|l}
Particle & $\tilde{f}$ & $\tilde{x}$ & $\tilde{b}$ & Touch & Rigidity & Range\\
& & & & $z/2 \le \tilde{x}$ & $z \tilde{f} /2 \ge \tilde{b}$ & \\
\hline
Frictionless Sphere & 1 &$d$ &$d$ & $z \le 2d$ & $z \ge 2d$ & $z = 2d$ \\
Frictional Sphere & $d$ &$d$ &$d(d\!+\!1)/2$& $z \le 2d$ & $z \ge d\!+\!1$ &  $d\!+\!1 \le z \le 2d$ \\
Frictionless Spheroid &1 &5 &5 & $z \le 10 $ & $z \ge 10$ &  $ z = 10$ \\
Frictional Spheroid &$3$ &5& $%d(d\!+\!1)/2 =
6$& $z \le 10 $ & $z \ge %d\!+\!1=
4$  &  $4 \le z \le 10$ \\
Frictionless Ellipsoid &1 &$d(d\!+\!1)/2$ &$d(d\!+\!1)/2$ & $z \le d(d\!+\!1)$ & $z \ge $ $d(d\!+\!1)$ &  $z =d(d\!+\!1)$ \\
Frictional Ellipsoid & d &$d(d\!+\!1)/2$ &$d(d\!+\!1)/2$ &$z \le d(d\!+\!1)$ & $z \ge $ $d\!+\!1$&  $d\!+\!1 \le z \le d(d\!+\!1)$ \\

\end{tabular}
\caption{\label{table:count} Results of ``Maxwell'' constraint
counting for a range of different type of soft particles. As
explained in the text, $\tilde{f}$ denotes the number of force
components per contact, $\tilde{x}$ denotes the geometrical number
of freedom per particle and $\tilde{b}$ denotes the number of
balance equations per particle.}
%\begin{indented}
%\end{indented}
\end{table*}

{\bf Absence of Floppy Modes ---} The counting that follows from
requiring that there are no floppy modes can most easily be
carried out by considering $\Delta E$, the change in elastic
energy as function of deformation of a certain packing. The number
of terms contributing to $\Delta E$ equals the number of contacts,
$N z/2$, multiplied with $\tilde{f}$, the number of independent
degrees of freedom necessarily to describe changes in the energy
at a contact. $\Delta E$ is a function of all $N d$ positional
degrees of freedom, and all additional orientational degrees of
freedom which are not symmetries --- zero for spheres, $2N$ for
spheroids in three dimensions, and $d(d-1)N/2$ for general
ellipsoids. We denote these number of degrees of freedom relevant
for $\Delta E$ by $\tilde{b}$.

Absence of generic floppy modes requires that the number of terms
contributing to $\Delta E$ exceeds the number of degrees of
freedom: $z \tilde{f}/2 \ge \tilde{b}$.

For frictionless particles, $\tilde{f}$ equals one, because energy
changes result from (de)compression of contacts only, while for
frictional particles, $\tilde{f}$ equals $d$, since relative
motion of contacting particles in all directions are relevant.

The situation for $\tilde{b}$ is simple for frictional particles,
where all positional and orientational degrees of freedom are
relevant and $\tilde{b}= d(d+1)/2$. For frictionless particles,
$\tilde{b}$ depends on the symmetries. For frictionless spheres,
only translational degrees of freedom are important and $\tilde{b}
= d$. For frictionless spheroids in three dimensions, two
additional rotational degrees of freedom come into play and
$\tilde{b}=5$, while for general frictionless ellipsoids, all
rotational degrees are relevant and $\tilde{b}=d(d+1)/2$.

{\bf Equivalence of Floppy Mode and Force Balance Counting ---}
The requirement $z \tilde{f}/2 \ge \tilde{b}$ is exactly the same
as requiring that there are sufficient contact forces in the
system so that they generically can be expected to balance: the
number of contact force degrees of freedom per particles is $z
\tilde{f}/2$ and the number of equations that need to be satisfied
equals $\tilde{b}$. The number of relevant particle degrees of
freedom in the energy expansion thus corresponds to the number of
force balance equations, and the number of terms in $\Delta E$ (=
number of constraints needed to generically avoid floppiness)
correspond to the number of force degrees of freedom
--- changes in energy and forces are directly linked.

Note that even though the role of constraints and degrees of
freedom interchanges when altering the picture between absence of
floppy modes and satisfaction of force balance, so does the
requirement (floppy modes: making sure there are no generic
solutions, force balance: making sure there are generic solution),
and in the force balance picture one ends up with precisely the
same inequality: $z \tilde{f}/2 \ge \tilde{b}$.

{\bf Touch ---} The conditions that particles precisely touch
yields $N z/2$ constraints on the degrees of freedom of the
particles. Denoting the number of geometric degrees per particle
as $\tilde{x}$, the condition that for generic packings there
should be less constraints than degrees of freedom yields $z/2 \le
\tilde{x}$.

For the particles that are considered here (spheres and ellipsoids
with and without friction), the number of degrees of freedom per
particles are their $d$ positional coordinates, to which
ellipsoids add their relevant angular degrees of freedom. For
general ellipsoids, these yield $d(d+1)/2$ degrees of freedom
--- for spheroids in three
dimensions (an ellipsoid with two equal axes, which thus has one
symmetry of rotation --- see section~\ref{sec:ellipsoid}) these
yield $5$ degrees of freedom. The resulting counting of
$\tilde{x}$ and corresponding inequalities are listed in table
\ref{table:count}. In particular, for frictional particles, the
lower bound for the contact number is $d+1$, while for
frictionless particles it depends on the symmetries of the
particles.

{\bf Results ---} The resulting inequalities are listed in table
\ref{table:count}. Note that the upper bounds for $z$ coincide for
frictional and frictionless particles, as this number only depends
on the geometrical number of degrees of freedom. The inequalities
can be summarized as follows: For frictionless particles,
$\tilde{f}$ equals one, $ \tilde{b} = \tilde{f}$ and the lower and
upper bounds coincide at $z=2\tilde{x} = 2 \tilde{b}/\tilde{f}$.
For frictional particles, $2\tilde{x}
> 2 \tilde{b}/\tilde{f}$, the lower and upper bounds do not
coincide, and a range of contact numbers is allowed at jamming.


\begin{thebibliography}{100}
\bibitem{philipseprivcom} The data also suggested that
$z$ decays monotonically for large $a$ but this is likely due to a
problem with the contact counting (A P Philipse, Priv. Comm.)
\bibitem{p16} A trivial scaling of $\omega \sim P^{1/6}$,
characteristic for Hertzian contacts, has been scaled out.
\bibitem{p13} A trivial
scaling of $K,G \sim P^{1/3}$, characteristic for Hertzian
contacts, has been scaled out
\end{thebibliography}
\end{document}